\documentclass[twocolumn,trackchanges]{aastex701}
\usepackage{booktabs}
\usepackage{amsmath} 

\begin{document}

\title{ASTROFLOW: A Real-Time End-to-End Pipeline for Radio Single-Pulse Searches}

\author[orcid=0009-0006-0364-4161]{Guanhong Lin}
\affiliation{National Astronomical Observatories, Chinese Academy of Sciences, 100101 Beijing, China; \href{mailto:jlzhang@nao.cas.cn}{jlzhang@nao.cas.cn}, \href{mailto:duanran@nao.cas.cn}{duanran@nao.cas.cn}}
\affiliation{University of Chinese Academy of Sciences, Beijing 100049, People’s Republic of China}
\email{lingh@bao.ac.cn}

\author{Dejia Zhou}
\affiliation{National Astronomical Observatories, Chinese Academy of Sciences, 100101 Beijing, China; \href{mailto:jlzhang@nao.cas.cn}{jlzhang@nao.cas.cn}, \href{mailto:duanran@nao.cas.cn}{duanran@nao.cas.cn}}
\affiliation{University of Chinese Academy of Sciences, Beijing 100049, People’s Republic of China}
\email{lingh@bao.ac.cn}

\author[orcid=0000-0002-2940-4821]{Jianli Zhang}
\affiliation{National Astronomical Observatories, Chinese Academy of Sciences, 100101 Beijing, China; \href{mailto:jlzhang@nao.cas.cn}{jlzhang@nao.cas.cn}, \href{mailto:duanran@nao.cas.cn}{duanran@nao.cas.cn}}
\email{lingh@bao.ac.cn}

\author{Jialang Ding}
\affiliation{National Astronomical Observatories, Chinese Academy of Sciences, 100101 Beijing, China; \href{mailto:jlzhang@nao.cas.cn}{jlzhang@nao.cas.cn}, \href{mailto:duanran@nao.cas.cn}{duanran@nao.cas.cn}}
\affiliation{University of Chinese Academy of Sciences, Beijing 100049, People’s Republic of China}
\email{lingh@bao.ac.cn}

\author{Fei Liu}
\affiliation{National Astronomical Observatories, Chinese Academy of Sciences, 100101 Beijing, China; \href{mailto:jlzhang@nao.cas.cn}{jlzhang@nao.cas.cn}, \href{mailto:duanran@nao.cas.cn}{duanran@nao.cas.cn}}
\email{lingh@bao.ac.cn}

\author{Xiaoyun Ma}
\affiliation{National Astronomical Observatories, Chinese Academy of Sciences, 100101 Beijing, China; \href{mailto:jlzhang@nao.cas.cn}{jlzhang@nao.cas.cn}, \href{mailto:duanran@nao.cas.cn}{duanran@nao.cas.cn}}
\email{lingh@bao.ac.cn}

\author{Yuan Liang}
\affiliation{National Astronomical Observatories, Chinese Academy of Sciences, 100101 Beijing, China; \href{mailto:jlzhang@nao.cas.cn}{jlzhang@nao.cas.cn}, \href{mailto:duanran@nao.cas.cn}{duanran@nao.cas.cn}}
\affiliation{University of Chinese Academy of Sciences, Beijing 100049, People’s Republic of China}
\email{lingh@bao.ac.cn}

\author{Ruan Duan}
\affiliation{National Astronomical Observatories, Chinese Academy of Sciences, 100101 Beijing, China; \href{mailto:jlzhang@nao.cas.cn}{jlzhang@nao.cas.cn}, \href{mailto:duanran@nao.cas.cn}{duanran@nao.cas.cn}}
\email{lingh@bao.ac.cn}

\author{Liaoyuan Liu}
\affiliation{National Astronomical Observatories, Chinese Academy of Sciences, 100101 Beijing, China; \href{mailto:jlzhang@nao.cas.cn}{jlzhang@nao.cas.cn}, \href{mailto:duanran@nao.cas.cn}{duanran@nao.cas.cn}}
\affiliation{University of Chinese Academy of Sciences, Beijing 100049, People’s Republic of China}
\email{lingh@bao.ac.cn}

\author{Xuanyu Wang}
\affiliation{National Astronomical Observatories, Chinese Academy of Sciences, 100101 Beijing, China; \href{mailto:jlzhang@nao.cas.cn}{jlzhang@nao.cas.cn}, \href{mailto:duanran@nao.cas.cn}{duanran@nao.cas.cn}}
\affiliation{University of Chinese Academy of Sciences, Beijing 100049, People’s Republic of China}
\email{lingh@bao.ac.cn}

\author{Xiaohui Yan}
\affiliation{National Astronomical Observatories, Chinese Academy of Sciences, 100101 Beijing, China; \href{mailto:jlzhang@nao.cas.cn}{jlzhang@nao.cas.cn}, \href{mailto:duanran@nao.cas.cn}{duanran@nao.cas.cn}}
\affiliation{University of Chinese Academy of Sciences, Beijing 100049, People’s Republic of China}
\email{lingh@bao.ac.cn}

\author{Yingrou Zhan}
\affiliation{National Astronomical Observatories, Chinese Academy of Sciences, 100101 Beijing, China; \href{mailto:jlzhang@nao.cas.cn}{jlzhang@nao.cas.cn}, \href{mailto:duanran@nao.cas.cn}{duanran@nao.cas.cn}}
\affiliation{University of Chinese Academy of Sciences, Beijing 100049, People’s Republic of China}
\email{lingh@bao.ac.cn}

\author{Yuting Chu}
\affiliation{National Astronomical Observatories, Chinese Academy of Sciences, 100101 Beijing, China; \href{mailto:jlzhang@nao.cas.cn}{jlzhang@nao.cas.cn}, \href{mailto:duanran@nao.cas.cn}{duanran@nao.cas.cn}}
\affiliation{University of Chinese Academy of Sciences, Beijing 100049, People’s Republic of China}
\email{lingh@bao.ac.cn}

\author{Jing Qiao}
\affiliation{National Astronomical Observatories, Chinese Academy of Sciences, 100101 Beijing, China; \href{mailto:jlzhang@nao.cas.cn}{jlzhang@nao.cas.cn}, \href{mailto:duanran@nao.cas.cn}{duanran@nao.cas.cn}}
\affiliation{University of Chinese Academy of Sciences, Beijing 100049, People’s Republic of China}
\email{lingh@bao.ac.cn}

\author{Wei Wang}
\affiliation{National Astronomical Observatories, Chinese Academy of Sciences, 100101 Beijing, China; \href{mailto:jlzhang@nao.cas.cn}{jlzhang@nao.cas.cn}, \href{mailto:duanran@nao.cas.cn}{duanran@nao.cas.cn}}
\email{lingh@bao.ac.cn}

\author{Jie Zhang}
\affiliation{School of Arts and Sciences, Shanghai Dianji University,Shanghai 200240, China; \href{mailto:zhangjie\_mail@126.com }{zhangjie\_mail@126.com}}
\affiliation{College of Physics and Electronic Engineering, Qilu Normal University, 2 Wenbo Road, Jinan 250300, China;\href{mailto:liumeng35@qlnu.edu.cn }{liumeng35@qlnu.edu.cn}}
\email{lingh@bao.ac.cn}

\author{Zerui Wang}
\affiliation{College of Physics and Electronic Engineering, Qilu Normal University, 2 Wenbo Road, Jinan 250300, China;\href{mailto:liumeng35@qlnu.edu.cn }{liumeng35@qlnu.edu.cn}}
\email{lingh@bao.ac.cn}

\author{Meng Liu}
\affiliation{College of Physics and Electronic Engineering, Qilu Normal University, 2 Wenbo Road, Jinan 250300, China;\href{mailto:liumeng35@qlnu.edu.cn }{liumeng35@qlnu.edu.cn}}
\email{lingh@bao.ac.cn}

\author{Chenchen Miao}
\affiliation{College of Physics and Electronic Engineering, Qilu Normal University, 2 Wenbo Road, Jinan 250300, China;\href{mailto:liumeng35@qlnu.edu.cn }{liumeng35@qlnu.edu.cn}}
\email{lingh@bao.ac.cn}

\author{Menquan Liu}
\affiliation{School of Arts and Sciences, Shanghai Dianji University,Shanghai 200240, China; \href{mailto:zhangjie\_mail@126.com }{zhangjie\_mail@126.com}}
\affiliation{College of Physics and Electronic Engineering, Qilu Normal University, 2 Wenbo Road, Jinan 250300, China;\href{mailto:liumeng35@qlnu.edu.cn }{liumeng35@qlnu.edu.cn}}
\email{lingh@bao.ac.cn}

\author{Meng Guo}
\affiliation{National Supercomputing Center in Jinan, No. 28666 Jing Shi East Road, Jinan, 250103, China}
\email{lingh@bao.ac.cn}

\author{Di Li}
\affiliation{Department of Astronomy, Tsinghua University, Beijing 100084, China}
\affiliation{National Astronomical Observatories, Chinese Academy of Sciences, Beijing 100101, China}
\affiliation{Research Center for Astronomical Computing, Zhejiang Laboratory, Hangzhou, 311100, China}
\affiliation{New Cornerstone Science Laboratory, Shenzhen, 518054, China}
\email{lingh@bao.ac.cn}

\author{Pei Wang}
\affiliation{National Astronomical Observatories, Chinese Academy of Sciences, 100101 Beijing, China; \href{mailto:jlzhang@nao.cas.cn}{jlzhang@nao.cas.cn}, \href{mailto:duanran@nao.cas.cn}{duanran@nao.cas.cn}}
\email{lingh@bao.ac.cn}








\begin{abstract}

Fast radio bursts (FRBs) are extremely bright, millisecond-duration cosmic transients of unknown origin. The growing number of wide-field and high–time-resolution radio surveys, particularly with next-generation facilities such as the SKA and MeerKAT, will dramatically increase FRB discovery rates, but also produce data volumes that overwhelm conventional search pipelines.  Real-time detection thus demands software that is both algorithmically robust and computationally efficient.  
We present \textsc{Astroflow}, an end-to-end, GPU-accelerated pipeline for single-pulse detection in radio time–frequency data. Built on a unified C++/CUDA core with a Python interface, Astroflow integrates RFI excision, incoherent dedispersion, dynamic-spectrum tiling, and a YOLO-based deep detector. Through vectorized memory access, shared-memory tiling, and OpenMP parallelism, it achieves $>10\times$ faster-than-real-time processing on consumer GPUs for a typical 150 s, 2048-channel observation—while preserving high sensitivity across a wide range of pulse widths and dispersion measures.
These results establish the feasibility of a fully integrated, GPU–accelerated single–pulse search stack, capable of scaling to the data volumes expected from upcoming large–scale surveys. \textsc{Astroflow} offers a reusable and deployable solution for real–time transient discovery, and provides a framework that can be continuously refined with new data and models.

\end{abstract}



\section{introduction}

Fast radio bursts (FRBs; \citet{Lorimer2007}) are short-duration, bright radio pulses that exhibit pronounced dispersion signatures \citep{Chatterjee2017,Bannister2019b}. Since their first discovery by \citet{Lorimer2007}, FRBs have become a central focus of time-domain radio astronomy. Within the broader class of transients, rotating radio transients (RRATs) reported by \citet{mclaughlin_transient_2006} likewise underscore the importance of single-pulse searches. For single pulses from distant sources, propagation through the interstellar medium causes frequency-dependent time delays. The most prominent is the dispersion delay from the cold ionized medium, whose magnitude is defined by the dispersion measure (DM)—the line-of-sight integral of the free-electron number density.

Single-pulse searches are typically conducted on data with high time and frequency resolution. A standard workflow comprises: (i) dedispersing the data over a series of trial dispersion measures (DMs; recent multicore and GPU–accelerated methods have enabled efficient implementations of this step \citep{acceleratingded-2023}); (ii) summing frequency channels to form a time series; and (iii) matched–filtering the time series with a bank of boxcar windows at multiple widths to enhance sensitivity to pulses of different durations. 

Building on this approach, a range of software packages and pipelines has been developed: \texttt{PRESTO} \citep{presto2011}, \texttt{HEIMDALL} \citep{Barsdell2012,Barsdell2024}, \texttt{PREDADA} \citep{BannisterFREDDA2019}, \texttt{AMBER} \citep{AMBER2020}, and \texttt{ASTRO-ACCELERATE} \citep{KarelSNR2020}. The CPU–oriented \texttt{PRESTO} decouples RFI excision, dedispersion, and single–pulse search modules \citep{presto2011}, whereas \texttt{HEIMDALL} is a widely used GPU–accelerated pipeline \citep{Barsdell2012}. Both have been applied in multiple observing projects (e.g., Tianlai \citep{yu2024frbsearchingpipelinetianlaicylinder}; GBT \citep{gbtrans_2019}). In addition, there exist end-to-end, CPU-based, highly concurrent single-pulse techniques (e.g., \texttt{transientX}; \citealt{transientx2024}).

Despite the maturity of current workflows, two practical challenges remain. First, real telescope backgrounds are dominated by coloured, often non-stationary noise rather than ideal Gaussian statistics\citep{Zhang2021coloured}, and they coexist with complex radio-frequency interference (RFI); together they can dramatically inflate the number of candidates—especially in wide parameter-space searches—leading to tens of thousands of triggers per day and substantial manual-vetting costs. Second, fixed S/N thresholds can discard near-threshold but genuine pulses, thereby compromising completeness. 

To address these issues, improvements have followed two lines. On the classical side, more robust RFI excision and quality control—such as filtering signals with $\mathrm{DM}\!\approx\!0$, adaptive bandpass suppression, and wavelet-domain interference mitigation—are used to reduce false positives while recovering S/N \citep{zerodmfilter2009,Dumez-Viou2016,wangpei2025}. On the machine-learning side, deep learning has been employed for backend binary classification of candidates, reducing human effort while maintaining recall \citep{Connor2018,Agarwal2020}. Recent work further explores applying deep learning directly to front-end single-pulse detection: for example, \textsc{DRAFTS} adopts an object-detection paradigm on dedispersed data to balance recall and false-positive control \citep{DRAFTS2025}; other studies attempt de-dispersion–free detection on raw spectrograms using attention mechanisms, enabling end-to-end modeling \citep{Xuerong2025}.

Looking ahead, next-generation facilities such as the SKA \citep{Dewdney2009} will deliver higher time resolution, finer channelization, and multi-beam observations, steadily increasing transient data rates and stressing traditional workflows. In this setting, it becomes increasingly important for single-pulse searches to sustain \emph{faster-than-real-time} throughput and to improve completeness—recovering more genuine pulses without a corresponding rise in false positives—within a cohesive software stack that mitigates module fragmentation and remains straightforward to deploy across heterogeneous backends.

This paper presents \textsc{astroflow}, an end-to-end single-pulse search software system. The design emphasizes algorithm–engineering integration, comprising: multi-format I/O with stable RFI-excision modules; a system-optimized, GPU-accelerated data-processing and dedispersion implementation; and fast candidate detection based on \texttt{YOLO}, accompanied by visualized and parameterized candidate outputs. At the engineering level, \textsc{Astroflow} leverages OpenMP, loop unrolling, and vectorization to reduce end-to-end latency and increase throughput. On representative time-domain datasets, the system achieves processing speeds significantly faster than real time while maintaining high detection rates with controllable false positives.

The remainder of this paper is organized as follows: Section~2 provides a system overview and key algorithmic implementations; Section~3 presents performance comparisons on standard datasets and mainstream software packages; Section~4evaluates \textsc{Astroflow} on the \texttt{FAST\_FREX} corpus and on a four-antenna 4.5,m L-band array; we validate with Crab giant pulses and report a ten-day continuous FRB survey; Section~5 discusses limitations and potential improvements; and Section~6 concludes the paper.

\section{System Overview}

\begin{figure*}
    \centering
    \includegraphics[width=0.95\linewidth]{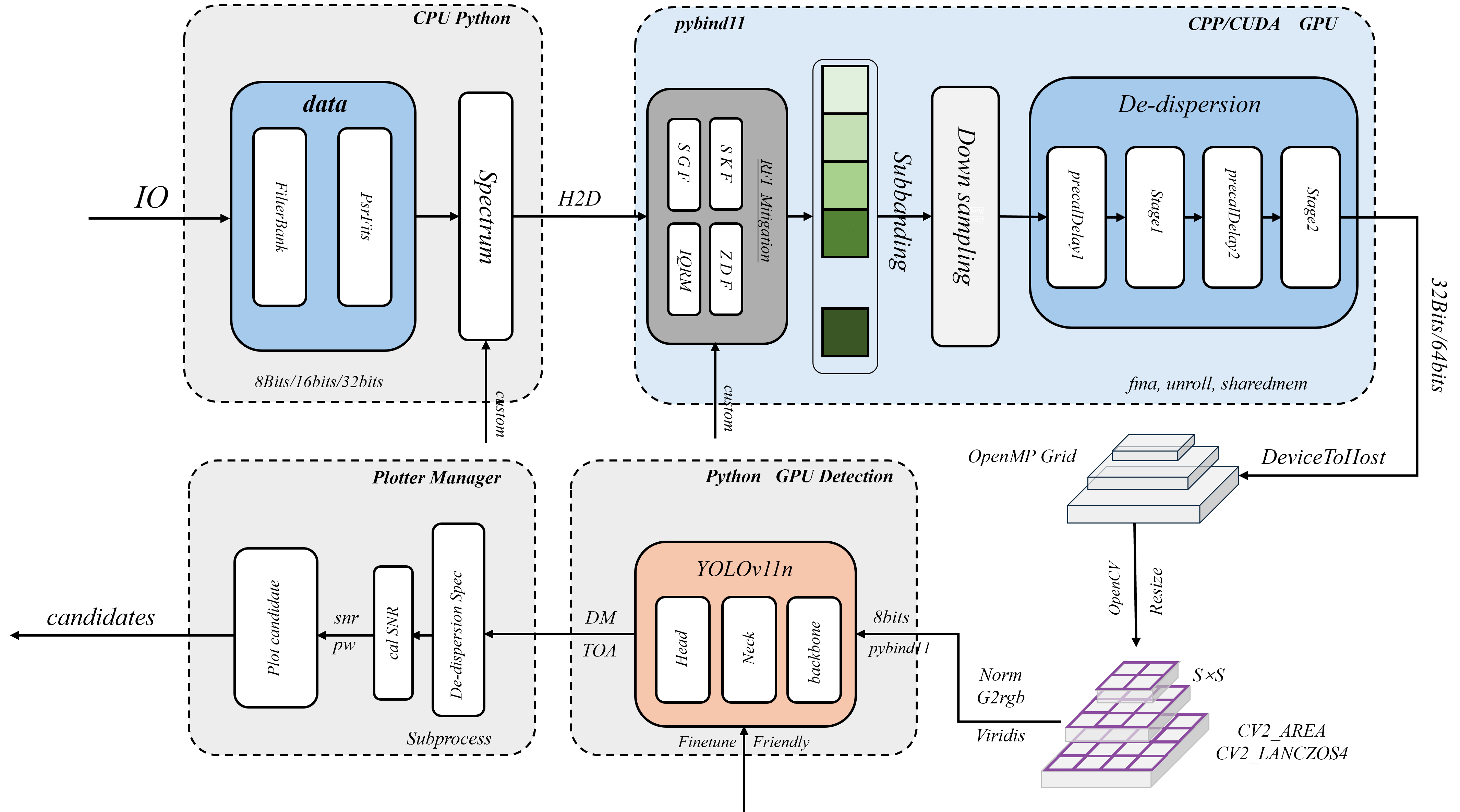}
    \caption{\textbf{Block diagram of the \textsc{Astroflow} pipeline.}
    Python-side ingestion (\texttt{Filterbank}, \texttt{PSRFITS}) bridges via \texttt{pybind11} to a C++/CUDA backend for RFI excision, subbanding, optional downsampling, and incoherent dedispersion; products feed a \texttt{YOLOv11n} detector and a plotter/producer. Arrows indicate data flow; gray denotes Python modules, blue denotes C++/CUDA kernels, and orange denotes the neural-network component.} 
    \label{fig:astroflow-pipe}
\end{figure*}

\textsc{Astroflow} integrates algorithmic modules with an engineered GPU pipeline that is accessed from Python. Figure~\ref{fig:astroflow-pipe} sketches the end-to-end data path and the major components.

At ingress, the system supports two radio-astronomy formats—\texttt{Filterbank} and \texttt{PSRFITS}—with extensible adapters for high-bandwidth storage. Dynamic spectra are decoded and organized in host memory, then moved to device memory via host-to-device (H2D) transfers. On the GPU, radio-frequency–interference (RFI) excision operates primarily along the frequency axis; configurable algorithms include the zero-DM filter \citep{zerodmfilter2009}, IQRM \citep{Morello2022}, a spectral–kurtosis filter \citep{Nita2016}, and Savitzky–Golay filter \citep{Agarwal2022sg}. Frequency subbanding and optional time downsampling follow, jointly controlling computational cost and improving the visibility of broader pulses.

The core compute stage performs parallel-optimized incoherent dedispersion, producing \( \mathrm{DM}\text{--}t \) slices. Kernels are tuned with fused multiply–add (FMA), loop unrolling, and shared-memory tiling. The dedispersed products are returned to the host via device-to-host (D2H) transfers. After gridding, intensity matrices are converted to RGB images and globally normalized using OpenCV. Through \texttt{pybind11}, the data are zero-copy mapped to NumPy for seamless handoff to the Python detection stack.

Detection is performed by a lightweight \texttt{YOLOv11n} model in streaming mode, which reports per-frame candidate locations and confidences \citep{yolov11n2024}. For candidates passing S/N and DM criteria, the pipeline records the estimated DM and time of arrival (TOA; referenced to the top of the band), triggers a candidate–product routine to re-dedisperse within a local time window, measures S/N and pulse width, and emits standardized diagnostic figures for human vetting. When needed, outputs can be exported in YOLO format to support subsequent retraining or fine-tuning.

\textsc{Astroflow} is easily deployed from the Python Package Index and provides a lightweight command–line interface with YAML–based configuration, enabling straightforward integration into standard analysis workflows\footnote{\url{https://github.com/lintian233/astroflow}};

\subsection{Radio-Frequency Interference Mitigation}
\begin{figure*}[htbp]
    \centering
    \begin{minipage}[b]{0.48\linewidth}
        \centering
        \includegraphics[width=\linewidth]{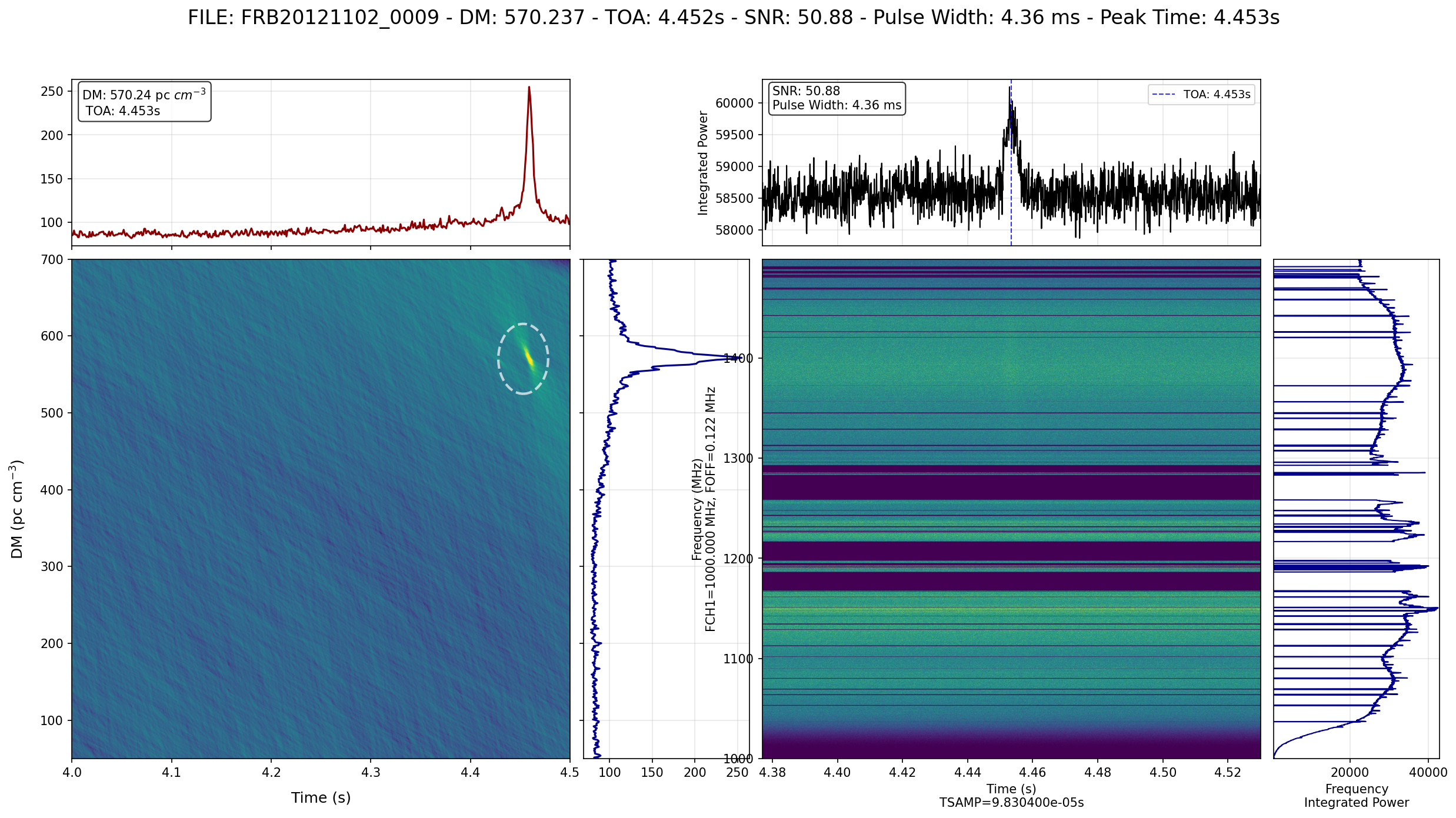}
    \end{minipage}
    \hfill
    \begin{minipage}[b]{0.48\linewidth}
        \centering
        \includegraphics[width=\linewidth]{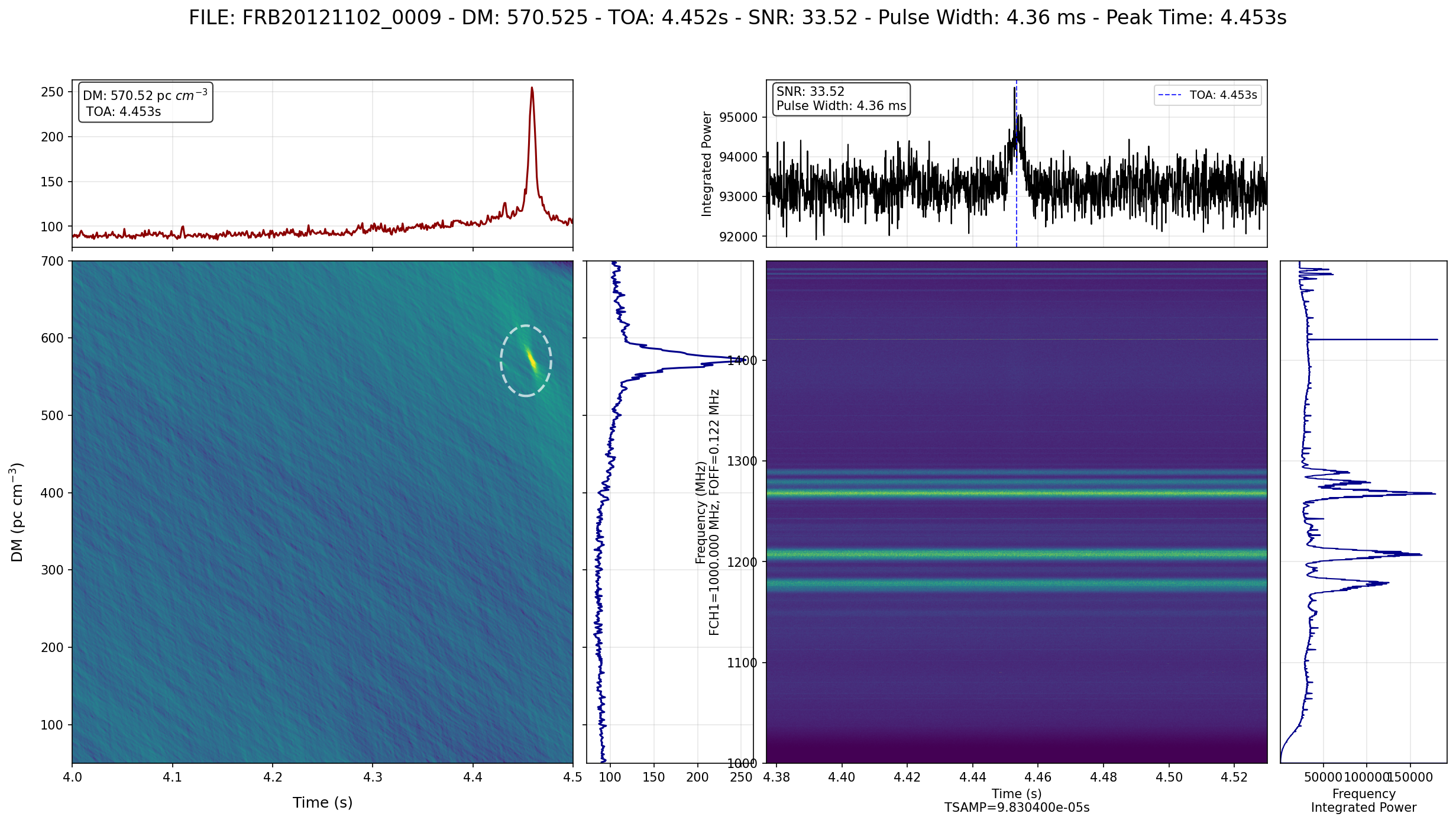}
    \end{minipage}
    \caption{Comparison of dynamic spectra before and after RFI mitigation.}
    \label{fig:rfi-comparison}
\end{figure*}

Radio–frequency interference (RFI), arising from anthropogenic electromagnetic emissions (e.g., electronic equipment and satellite transmissions), can produce numerous false positives in single–pulse searches. Common manifestations include narrowband and broadband interference. To address these contaminants, \textsc{astroflow} implements GPU versions of the \texttt{zero-DM filter} and \texttt{IQRM}(\citealt{zerodmfilter2009}; \citealt{Morello2022}), and integrates filters such as Savitzky–Golay filter\citep{Agarwal2022sg} and spectral kurtosis filter\citep{Nita2016}; these can be adaptively combined according to the RFI characteristics of a dataset. Figure~\ref{fig:rfi-comparison} illustrates the effect of RFI mitigation on the dynamic spectrum.

\subsection{Downsampling and Subband Combination}
In single–pulse searches, \textsc{Astroflow} can optionally apply time downsampling and frequency subbanding/combination to the dynamic spectrum. Time downsampling reduces computational load and can improve the visibility of broader pulses while suppressing very narrow RFI spikes; however, excessive downsampling may smear signals whose intrinsic widths approach the time resolution. To balance efficiency and fidelity, we adopt a square–law (square–root–compressed) downsampling strategy, which tends to preserve the prominence of transient peaks better than simple averaging. In addition, a single pulse does not always span the full observing band. In such cases, dividing the band into several subbands and combining/detecting them separately can improve visibility when parts of the band are RFI–contaminated or when the signal exhibits strong frequency dependence.

For a window of $N$ samples with non-negative intensities $\{I_k\}_{k=1}^{N}$ at a fixed channel, the value written to the downsampled stream is
\begin{equation}
\tilde{I} \;=\; \min\!\left\{\, V_{\max}\,\frac{\sqrt{\sum_{k=1}^{N} I_k}}{\sqrt{N\,V_{\max}}}\,,\; V_{\max}\,\right\}
\end{equation}
Here $V_{\max}$ is the maximum representable value under the chosen bit width (e.g., $2^{8}-1$, $2^{16}-1$), and $N$ is the number of time samples in the bin. The $\min$ enforces saturation at $V_{\max}$.

\subsection{Dedispersion}\label{sec:dedispersion}
When a radio wave propagates through a tenuous ionized medium (cold–plasma approximation), the group velocity is frequency dependent, and low frequencies lag higher ones. The group–delay difference between two frequencies $f_1$ and $f_2$ is
\begin{equation}
\tau_d=\frac{e^2}{2\pi m_e c}\,\mathrm{DM}\!\left(\frac{1}{f_1^2}-\frac{1}{f_2^2}\right),
\label{eq:phys_delay}
\end{equation}
where $e$, $m_e$, and $c$ denote the elementary charge, the electron mass, and the speed of light in vacuum, respectively. The dispersion measure is defined as
\begin{equation}
\mathrm{DM}=\int n_e\,\mathrm{d}l ,
\label{eq:dm_def}
\end{equation}
i.e., the electron column density along the line of sight (pc\,cm$^{-3}$). In practice, the dedispersion delay is commonly written as
\begin{equation}
\Delta t(f,\mathrm{DM})=K_{\mathrm{DM}}\,\mathrm{DM}\!\left(\frac{1}{f^2}-\frac{1}{f_{\mathrm{ref}}^2}\right),
\label{eq:KDM}
\end{equation}
with $f$ and $f_{\mathrm{ref}}$ in MHz and $K_{\mathrm{DM}}=4.148808\times10^{3}\ {\rm ms\;MHz^2\;pc^{-1}\,cm^3}$. On a discrete dynamic spectrum, integer sample delays are obtained by rounding $\Delta t/t_{\rm samp}$; if time downsampling is used, then $t_{\rm samp}^{\downarrow}=D\,t_{\rm samp}$, where $D$ is the time downsampling factor.

In single–pulse searches, a standard practice is to dedisperse over a set of trial dispersion measures and then sum over frequency to enhance the pulse signal. The most direct implementation is brute–force dedispersion on a fine, fixed–step DM grid, resampling and summing the entire band and time series for each DM, with computational complexity $\mathcal{O}\!\bigl(N_{\mathrm{DM}}\,N_t\,N_{f}\bigr)$. To reduce the cost, more efficient algorithms have been developed: tree dedispersion \citep{Taylor1974} reduces the complexity to approximately $\mathcal{O}\!\bigl(N_{\mathrm{DM}}\,N_{t}\,log_2 N_{f}\bigr)$ via divide–and–conquer accumulation across frequency; the fast dispersion–measure transform (FDMT; \citealt{Zackay2017}) is, in idealized settings, nearly independent of $N_{\mathrm{DM}}$ with a similar overall order; another line of work is subband dedispersion \citep{Magro2011}, which approximates the dispersion phase within a subband by a representative frequency and performs coarse alignment followed by fine residual shifts \citep[see also][]{Barsdell2012}.

In this paper, we adopt the subband strategy proposed by \citealt{Magro2011} \citep[see also][]{Barsdell2012} and implement a two-stage accumulation on CUDA. All channels are first partitioned into subbands, each represented by a frequency (e.g., the median or a $1/f^2$–weighted center), and the trial–DM sequence is grouped with a coarse step. Let the fine step be $\Delta\mathrm{DM}$ and let an integer $N_0$ define the coarse step $\Delta\mathrm{DM}_0=N_0\Delta\mathrm{DM}$, with coarse grid $\mathrm{DM}_j=\mathrm{DM}_{\rm low}+j\,\Delta\mathrm{DM}_0$. After robust RFI mitigation  and time downsampling, the data enter two alignment–and–sum stages.

\emph{Stage~1 (channel level).} For each coarse DM, apply an integer delay and sum within each subband. For a channel center frequency $f_c$ relative to reference $f_{\rm ref}$, the integer delay on the downsampled time axis is
\begin{equation}
D_{j,c}=\mathrm{round}\!\left(\frac{K_{\mathrm{DM}}\,\mathrm{DM}_j}{t_{\rm samp}^{\downarrow}}
\left(\frac{1}{f_c^2}-\frac{1}{f_{\rm ref}^2}\right)\right).
\label{eq:coarse_delay}
\end{equation}
To avoid GPU–memory overflow and out–of–range indexing in Stage~2, we process the time axis in blocks: with block start $\tau_0$ and length $L$, set
\begin{equation}
L_1=\min\!\bigl(L+R_{\max},\,N_t^{\downarrow}-\tau_0\bigr),\qquad
R_{\max}=\max_{m,s}\bigl|R_{m,s}\bigr| ,
\label{eq:L1}
\end{equation}
where $R_{m,s}$ are the residual shifts defined below. If the $s$-th subband contains channels $\mathcal{C}_s$, the Stage~1 subband time series (zero outside bounds) is
\begin{equation}
I(j,s,u)=\sum_{c\in\mathcal{C}_s}X^{\downarrow}\!\bigl(c,\tau_0+u+D_{j,c}\bigr),
\qquad u=0,\ldots,L_1-1 .
\label{eq:stage1}
\end{equation}

\emph{Stage~2 (subband level).} For each fine DM $\mathrm{DM}_m$, let $j(m)=\lfloor m/N_0\rfloor$ be its coarse–group index and $\Delta\mathrm{DM}_m=\mathrm{DM}_m-\mathrm{DM}_{j(m)}$ the residual. Using the representative frequency $F_s$ of subband $s$, the residual shift on the downsampled grid is
\begin{equation}
R_{m,s}=\mathrm{round}\!\left(
\frac{K_{\mathrm{DM}}\,\Delta\mathrm{DM}_m}{t_{\rm samp}^{\downarrow}}
\left(\frac{1}{F_s^2}-\frac{1}{f_{\rm ref}^2}\right)\right),
\label{eq:residual}
\end{equation}
and, writing back only the first $L$ samples of the block, the dedispersed output for the fine DM is
\begin{equation}
\begin{aligned}
Y(\mathrm{DM}_m,\tau_0+v)=\sum_{s}I\!\bigl(j(m),s,v+R_{m,s}\bigr), \\
\qquad v=0,\ldots,L-1.
\label{eq:stage2}
\end{aligned}
\end{equation}
Concatenating the time blocks yields the full $N_{\mathrm{DM}}\times N_t^{\downarrow}$ dedispersed dynamic spectrum.

The subband approximation replaces the per–channel phase by that at $F_s$, thereby introducing additional broadening that grows with the subband width and the residual DM. Writing the subband width as $\Delta f_s$ around $F_s$ and neglecting intra–subband structure, a first–order estimate is
\begin{equation}
\begin{aligned}
\delta t_{\rm sb} &\simeq
\left|\frac{\partial}{\partial f}\!\left(K_{\mathrm{DM}}\,\Delta\mathrm{DM}\,\frac{1}{f^2}\right)\right|_{f=F_s} \Delta f_s
\\
&=\frac{2K_{\mathrm{DM}}\,|\Delta\mathrm{DM}|}{F_s^3}\,\Delta f_s .
\label{eq:sb_smear}
\end{aligned}
\end{equation}
Hence, increasing the number of subbands (smaller $\Delta f_s$), reducing the coarse step $N_0$ (smaller $|\Delta\mathrm{DM}|$), or adopting a phase–aware center (e.g., $1/f^2$ weighted) can effectively suppress this error. Compared with the naive $\mathcal{O}\!\bigl(N_{\mathrm{DM}}\,N_{f}\,N_{t}\bigr)$ scaling, the present method has time complexity
\begin{equation}
\mathcal{O}\!\bigl(N_{\mathrm{DM}}^{(\mathrm{coarse})}\,N_{f}\,N_{t}^{\downarrow}\bigr)
+\mathcal{O}\!\bigl(N_{\mathrm{DM}}\,N_{\mathrm{SB}}\,N_{t}^{\downarrow}\bigr),
\end{equation}
where $N_{\mathrm{DM}}^{(\mathrm{coarse})}=\lceil N_{\mathrm{DM}}/N_0\rceil$, $N_{\mathrm{SB}}\ll N_{f}$, and $N_{t}^{\downarrow}=\lceil N_t/D\rceil$.

In our implementation, we use time blocking and precompute both the per–channel coarse–DM delay table and the per–subband fine–DM residual–shift table, then execute the two–stage subband dedispersion. RFI mitigation is applied before downsampling. section \ref{sec:benchmarking} shows benchmarking results and a timing comparison against \texttt{Heimdall} \citep{Barsdell2012}.

\subsection{Griding and Rendering}\label{sec:viz}
After dedispersion, the dynamic spectrum forms a two–dimensional array
$X\in\mathbb{R}^{N_{\rm DM}\times N_{t}^{\downarrow}}$ on the ${\rm DM}$–time plane (row $i$ corresponds to the $i$-th trial ${\rm DM}$; column $k$ corresponds to the $k$-th downsampled time sample with spacing $t_{\rm samp}^{\downarrow}=D\,t_{\rm samp}$). For downstream automatic recognition, we slice the time axis into fixed-duration segments and resample each segment into an $M\times M$ pseudo–colour image. Let the slice duration be $T_{\rm slice}$; then the number of time samples per slice is
\begin{equation}
n_{\rm s}=\left\lceil \frac{T_{\rm slice}}{t_{\rm samp}^{\downarrow}}\right\rceil,\qquad 
S=\left\lceil \frac{N_{t}^{\downarrow}}{n_{\rm s}}\right\rceil ,
\end{equation}
yielding $S$ slices. The $s$-th slice ($s=0,\ldots,S-1$) corresponds to the column interval
$k\in\bigl[s\,n_{\rm s},\,\min\!\bigl((s+1)n_{\rm s},\,N_{t}^{\downarrow}\bigr)\bigr)$; denote the subarray by
$X^{(s)}\in\mathbb{R}^{N_{\rm DM}\times n_{\rm s}^{(s)}}$. We geometrically resample $X^{(s)}$ to
$U^{(s)}\in\mathbb{R}^{M\times M}$ with $M=512$. When both dimensions shrink
($N_{\rm DM}>M$ and $n_{\rm s}^{(s)}>M$), we use area–based anti–aliasing interpolation (equivalent to pixel–area aggregation); otherwise, we adopt Lanczos–4 interpolation to preserve narrow features\citep{opencv_library}. In all cases, columns map to time and rows to ${\rm DM}$ so that the aspect ratio of the time and ${\rm DM}$ axes is respected in visualization.

\begin{figure*}[htbp]
  \centering
  \setlength{\tabcolsep}{2pt} 
  \renewcommand{\arraystretch}{1.0}
  \begin{tabular}{cccc}
    \includegraphics[width=0.2\linewidth]{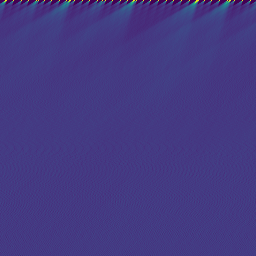} &
    \includegraphics[width=0.2\linewidth]{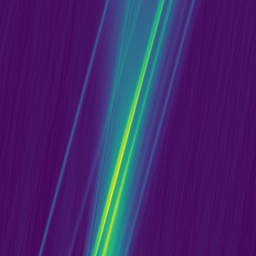} &
    \includegraphics[width=0.2\linewidth]{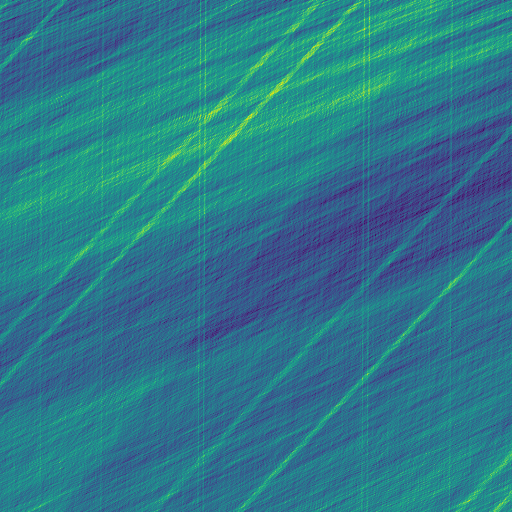} &
    \includegraphics[width=0.2\linewidth]{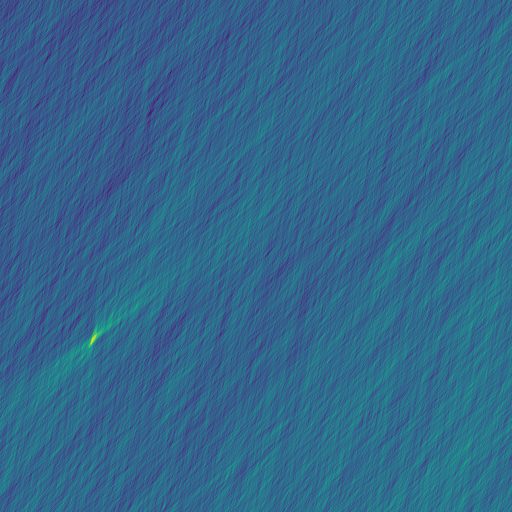} \\
    \includegraphics[width=0.2\linewidth]{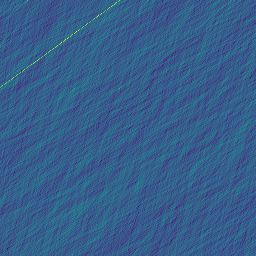} &
    \includegraphics[width=0.2\linewidth]{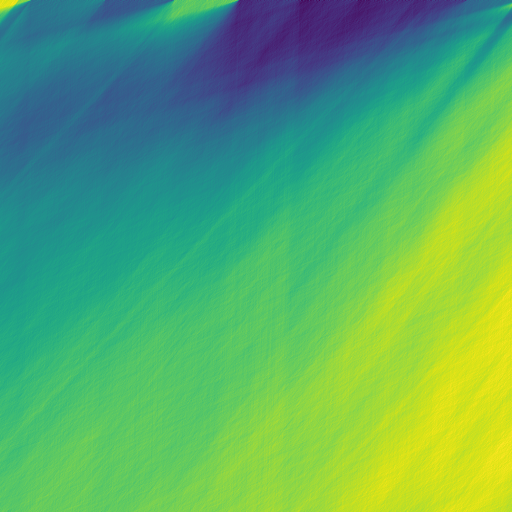} &
    \includegraphics[width=0.2\linewidth]{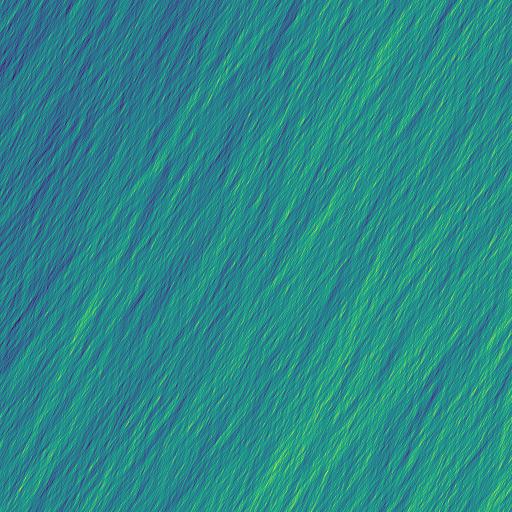} &
    \includegraphics[width=0.2\linewidth]{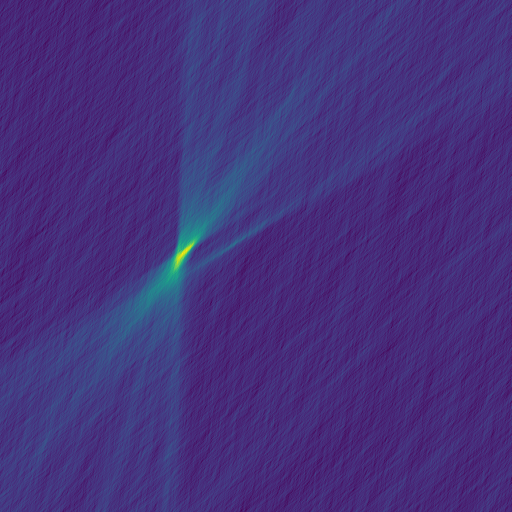}
  \end{tabular}
    \caption{\textbf{Examples of ${\rm DM}$–time tiles produced by the gridding and rendering.}
    All eight panels are $M\times M$ pseudo–colour images generated from the
    dedispersed data and rendered with a perceptually uniform colourmap.
    The first three columns show slices dominated by background and radio–frequency interference (RFI),
    whereas the two panels in the rightmost column exhibit the characteristic “bow–tie” single–pulse
    morphology—i.e., the specific targets of the downstream detection model.}
  \label{fig:dmt_example}
\end{figure*}

To remove inter–slice amplitude drift and enhance local contrast, each slice is independently linearly normalized and quantized to 8–bit grayscale. With $u_{\min}^{(s)}=\min U^{(s)}$ and $u_{\max}^{(s)}=\max U^{(s)}$, the normalized grayscale image is
\begin{equation}
\begin{aligned}
Y^{(s)}(i,j)=\mathrm{round}\!\left[
  255\,\frac{U^{(s)}(i,j)-u_{\min}^{(s)}}{u_{\max}^{(s)}-u_{\min}^{(s)}+\varepsilon}
\right], \\
\qquad Y^{(s)}\in\{0,\ldots,255\}^{M\times M},
\end{aligned}
\end{equation}
where $\varepsilon$ is a small positive constant for numerical stability. Finally, we apply a pseudo–colour mapping $\mathcal{C}$ (Viridis in this work) to obtain a three–channel colour image,
\begin{equation}
\mathcal{I}^{(s)}=\mathcal{C}\!\bigl(Y^{(s)}\bigr)\in[0,255]^{M\times M\times 3},
\end{equation}
using a perceptually uniform colourmap. Each time slice is thus represented as an $M\times M$ RGB image that serves as input to the deep–learning detection model. Figure~\ref{fig:dmt_example} shows an example ${\rm DM}$–time pseudo–colour image produced by this procedure.

\subsection{Candidate Visualization}
Once a candidate is detected by \texttt{YOLOv11n} \citep{Khanam2024}, its dispersion measure and time of arrival are obtained. An asynchronous plotting subprocess is then launched via \texttt{subprocess}. In this subprocess, the data segment around the candidate is dedispersed at the specified DM/TOA, and radio–frequency–interference (RFI) excision is applied to the corresponding dynamic spectrum. After dedispersion, post–processing computes the signal–to–noise ratio (S/N), pulse width, and a refined TOA. 

Three rendering modes are currently supported: (i) \texttt{std}, which outputs the raw dynamic spectrum; (ii) \texttt{subbands}, which visualizes a downsampled subband spectrum; and (iii) \texttt{detrend}, which applies normalization with detrending. Figure~\ref{fig:cand} shows an example of the candidate product. Depending on the configuration, the system packages the associated metadata and produces a finalized candidate image.

\subsection{Deep-Learning Detection Model}
With the rapid progress of machine learning—especially deep learning—an increasing number of studies have leveraged these methods to address limitations of traditional approaches. In radio astronomy, deep learning has been applied to radio–frequency–interference (RFI) mitigation \citep{Akeret2017} and radio image reconstruction \citep{Schmidt2022}, among other problems.

Two–dimensional ${\rm DM}$–time maps (and, alternatively, frequency–time spectrograms) are well suited to object–detection algorithms. Object detection is a central subfield of computer vision, with broad real–world applications such as video surveillance and autonomous driving; it aims to parse visual content by both recognizing object classes and localizing them within images. In recent years, advances in deep neural networks have markedly improved detector performance.

Within this landscape, \texttt{YOLOv11} (Ultralytics) continues the one–stage detector lineage, balancing high throughput and low latency with improved accuracy–parameter trade–offs. Relative to earlier generations, \texttt{YOLOv11} attains higher mAP on the COCO benchmark with fewer parameters, making it well suited to resource–constrained and real–time settings \citep{Khanam2024}. In particular, the nano–scale variant \texttt{YOLOv11n} has a parameter count on the order of $\sim$2.6M, which substantially reduces inference latency and memory usage while maintaining competitive detection quality\citep{Jocher_Ultralytics_YOLO_2023}.

In this work, we train a \texttt{YOLOv11N}–based single–pulse detector on ${\rm DM}$–time (or frequency–time) pseudo–colour slices: the network ingests $M\times M$ RGB inputs and outputs a bounding box (bbox) and confidence for each candidate, from which we infer the dispersion measure and time of arrival. The model training protocol and results are presented in Section~5.

\subsection{Candidate Plotting}
For each candidate detected by \texttt{YOLOv11n}, once the dispersion measure and time of arrival are determined, a plotting worker is launched asynchronously via a subprocess. Using the supplied DM/TOA, the data segment is re–dedispersed and subjected to radio–frequency–interference (RFI) mitigation. The dedispersed segment is then post–processed to produce the signal–to–noise ratio (S/N), pulse width, and a refined TOA.

\begin{figure}
    \centering
    \includegraphics[width=0.9\linewidth]{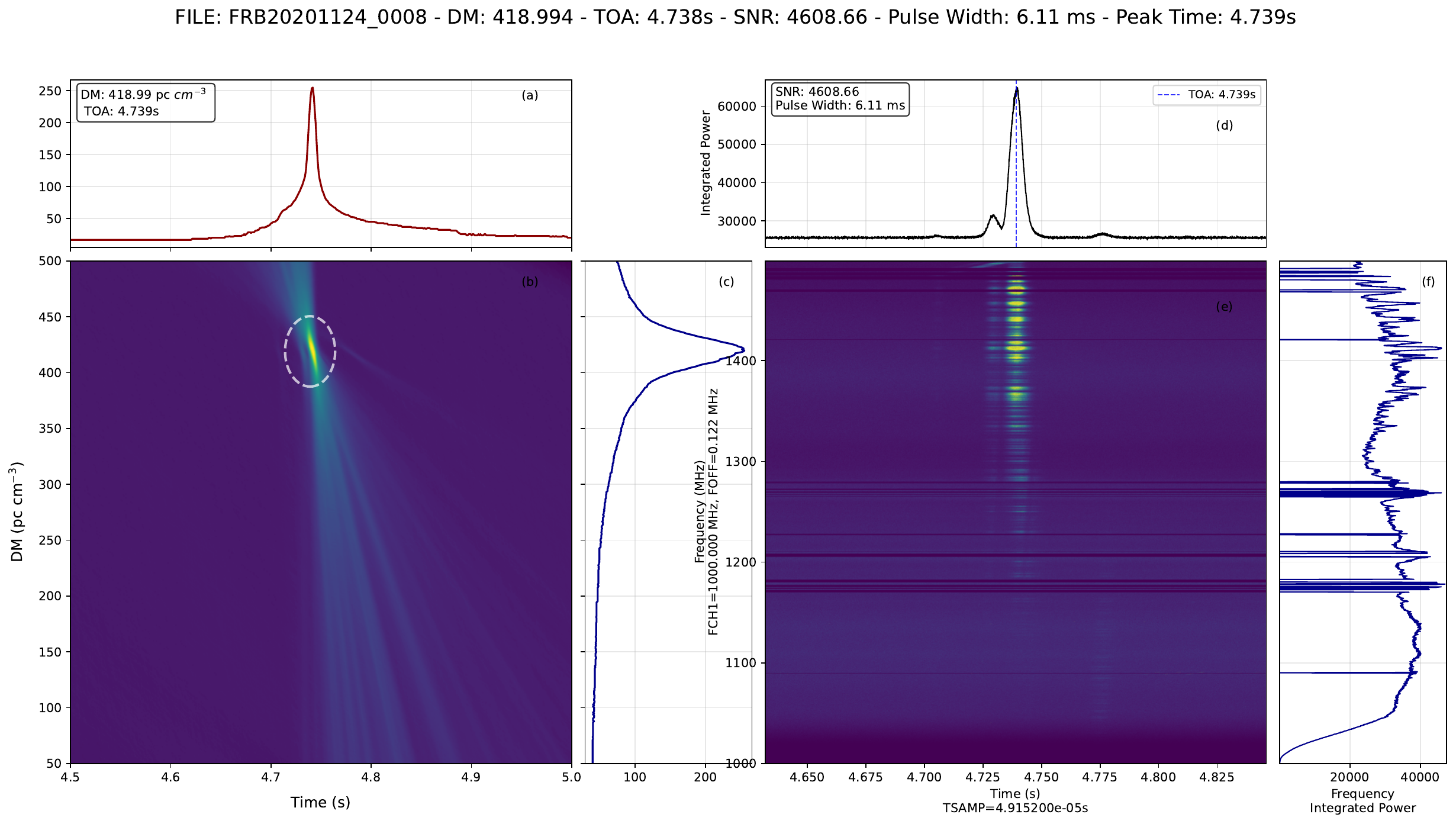}
    \caption{\textbf{Example candidate and panel layout.}
    The header lists the file identifier and post–processing measurements at the best–fit dispersion measure and time of arrival
    (\(\mathrm{DM}=418.994~\mathrm{pc\,cm^{-3}}\), \(\mathrm{TOA}=4.738~\mathrm{s}\)), along with the integrated signal–to–noise ratio
    (\(\mathrm{S/N}=4608.66\)) and pulse width (\(6.11~\mathrm{ms}\)).
    (a) Time series obtained by summing the \(\mathrm{DM}\)–time map in panel~(b) over \(\mathrm{DM}\) within the detection window.
    (b) \(\mathrm{DM}\)–time diagnostic after re–dedispersion; the dashed ellipse marks the detection window centered on the best
    \(\mathrm{DM}/\mathrm{TOA}\).
    (c) bandpass (integrated power as a function of radio frequency), annotated with the channelization parameters.
    (d) Dedispersed time series at the best \(\mathrm{DM}\); the measured \(\mathrm{TOA}\) is shown by the vertical dashed line and the
    \(\mathrm{S/N}\)/width are indicated in the legend.
    (e) Dedispersed frequency–time dynamic spectrum centered on the candidate; the burst appears as a near–vertical enhancement across the band.
    (f) Spectrum integrated over time (“frequency–integrated power”).}
    \label{fig:cand}
\end{figure}

At present, three rendering modes are supported: \emph{std}, which outputs the standard raw dynamic spectrum; \emph{subbands}, which displays a downsampled subband spectrum; and \emph{detrend}, which applies normalization with a de–trending strategy. Figure~\ref{fig:cand} shows an example candidate product. Depending on the configuration, the relevant metadata are packaged and a candidate–specific figure is generated and saved.

\section{Benchmarking}\label{sec:benchmarking}
In this section, we evaluate the end-to-end performance of \textsc{Astroflow}. We first examine how the GPU dedispersion runtime depends on input–data parameters. To compare our dedispersion with \texttt{Heimdall}\citep{Barsdell2012}, we consider two test cases. The first test (Section~3.1) fixes the dedispersion configuration and measures execution time as a function of the number of frequency channels. The second test (Section~3.2) evaluates performance versus the number of trial dispersion measures. Section~3.3 reports end-to-end timings on real observational data, including data ingestion, memory allocation, PCIe overhead, and per-stage timing throughout the pipeline. Owing to space constraints, we report 8-bit results only, although \textsc{Astroflow} supports 8/16/32-bit inputs.

We compare our results with widely used, HPC-oriented brute-force dedispersion implementations that are actively deployed. Specifically, for the dedispersion-focused tests we use the \texttt{dedisp} code \citep{Barsdell2012}. The \texttt{Heimdall} pipeline builds upon \texttt{dedisp} as its GPU engine. In its “adaptive” mode, the DM step varies according to a user-specified DM tolerance; by contrast, our implementation uses a fixed DM step and a fixed number of trials within each DM range. We ensure matched DM ranges and (nearly) matched trial counts across the comparisons. Our testbed consists of an 2 $\times$ Intel(R) Xeon(R) Silver 4314 CPU and an NVIDIA RTX~4090 GPU;

For timing, we use NVIDIA Compute Profiler (\texttt{ncu}) on the GPU side and 
\texttt{std::\allowbreak chrono::\allowbreak high\_resolution\_clock} on the CPU side. 
In addition to wall time, we report performance using the real-time factor.
\begin{equation}
    R \;=\; \frac{t_{\rm obs}}{t_c},
\end{equation}
where $t_{\rm obs}$ is the dedispersed observation span after truncation by the DM delay, and $t_c$ is the measured dedispersion compute time (kernel only), excluding host-to-device (H2D) and device-to-host (D2H) transfers.

\subsection{Frequency–Resolution Test for Dedispersion}
\begin{figure}
  \centering
  \includegraphics[width=0.95\linewidth]{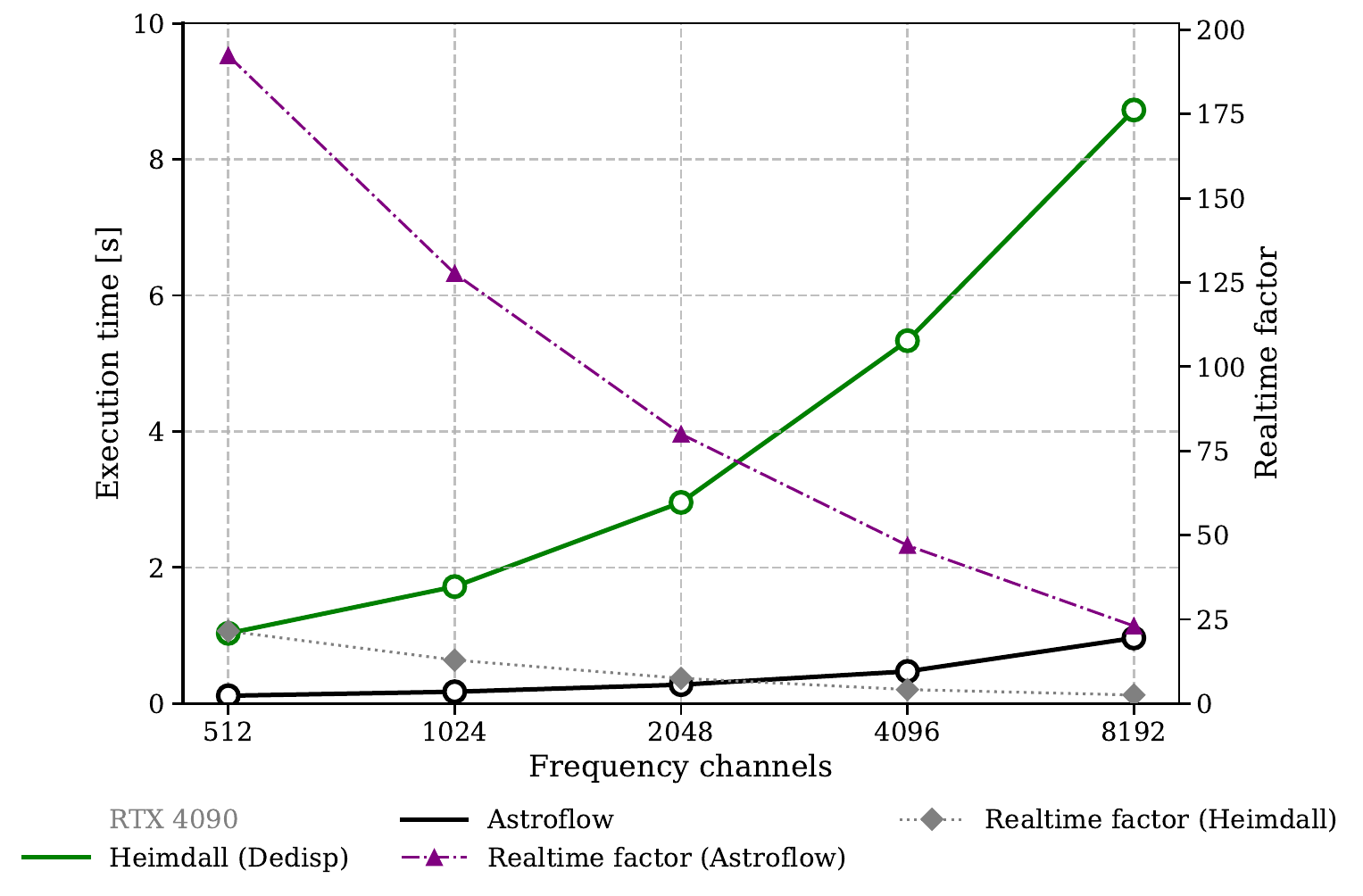}
  \caption{\textbf{Dedispersion runtime versus frequency–channel count.}
  Left axis: kernel execution time; right axis: real–time factor
  $R=t_{\rm obs}/t_c$.
  The benchmark uses a synthetic 8-bit dataset with duration
  $T=22\,\mathrm{s}$, sampling interval $t_s=40\,\mu\mathrm{s}$, center frequency
  $f_c=1250\,\mathrm{MHz}$, total bandwidth $500\,\mathrm{MHz}$, DM range $0$–$1024$,
  and $N_{\rm DM}=1500$ trials, executed on an RTX~4090.
  The green curve shows \texttt{Heimdall} (\texttt{dedisp});
  the black curve shows \textsc{Astroflow}. The purple dash–dot line (triangles)
  and the gray dotted line (diamonds) give $R$ for \textsc{Astroflow} and
  \texttt{Heimdall}, respectively. 
  As the number of channels increases from $512$ to $8192$, the
  \texttt{Heimdall} runtime grows from $\sim\!1.03$\,s to $\sim\!8.72$\,s
  ($R\!\approx\!22\rightarrow2$), while \textsc{Astroflow} remains markedly lower,
  from $\sim\!0.11$\,s to $\sim\!0.96$\,s ($R\!\approx\!190\rightarrow25$),
  maintaining faster–than–real–time performance across all tested channelizations.
  }
  \label{fig:dedisp-freqres}
\end{figure}

This test examines execution time as the number of frequency channels is varied across all codes under consideration. We generate a synthetic 8-bit dataset with a sampling interval of $40\,\mu{\rm s}$, a total duration of $22\,{\rm s}$, a center frequency of $1250\,{\rm MHz}$, and a total bandwidth of $500\,{\rm MHz}$. Five channelizations are tested: $512$, $1024$, $2048$, $4096$, and $8192$. The DM range is $0$ to $1024$ with $1500$ trial DMs. Because the maximum dispersion delay at ${\rm DM}=1024$ is $12.8\,{\rm s}$, unless otherwise stated the quoted duration refers to the \emph{dedispersion span} rather than the raw data length. A length of $22\,{\rm s}$ is sufficient to obtain representative timing results. For high-DM measurements, time downsampling is often applied; this particular test does not include downsampling. Figure~\ref{fig:dedisp-freqres} reports the timings versus the number of frequency channels.

\subsection{Dedispersion Timing versus Number of DM Trials}
In this subsection, we analyze the execution time of \textsc{Astroflow}'s dedispersion as a function of the number of trial dispersion measures ($N_{\rm DM}$) and compare it with the GPU–accelerated \texttt{Heimdall} pipeline, which is built on the \texttt{dedisp} library. The test configuration is: center frequency $f_c=1250$\,MHz, total bandwidth $500$\,MHz, DM range $0$–$1024$; for \textsc{Astroflow}, $N_{\rm DM}$ spans $512$–$8192$. For \texttt{Heimdall}, we use the “adaptive” mode, in which the DM step is determined at start-up from a user–specified broadening–tolerance factor (default 1.25), yielding a DM list with variable spacing \citep{Barsdell2012}. We choose tolerance values that produce trial counts approximately matching those of \textsc{Astroflow}. All runs use a dedispersion span of $20$\,s with 8-bit samples and a sampling interval of $40\,\mu$s.

\begin{figure}
  \centering
  \includegraphics[width=0.92\linewidth]{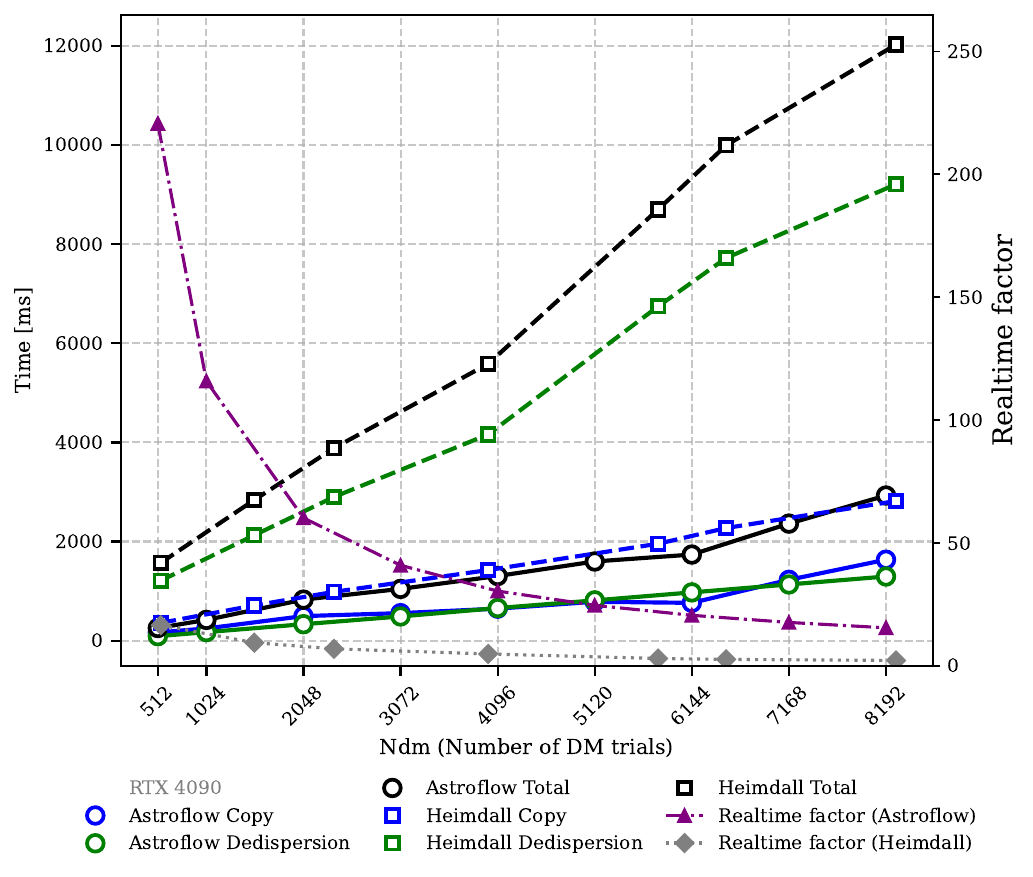}
  \caption{\textbf{Dedispersion timing versus the number of DM trials.}
  Test setup: 8-bit synthetic data with observation span $T=20$\,s, sampling interval $t_s=40\,\mu$s, 
  $n_f=2048$ channels, center frequency $f_c=1250$\,MHz, total bandwidth $500$\,MHz, and DM range $0$–$1024$;
  GPU: RTX~4090. Solid curves with circles show \textsc{Astroflow} times (Copy, Dedispersion, Total); 
  dashed curves with squares show \texttt{Heimdall} times (Copy, Dedispersion, Total), whose engine is 
  \texttt{dedisp}. The right axis gives the real–time factor $R=t_{\rm obs}/t_c$ computed 
  from the kernel time: purple dash–dot triangles for \textsc{Astroflow}, gray dotted diamonds for \texttt{Heimdall}.}
  \label{fig:astro-heim-timing}
\end{figure}

Figure~\ref{fig:astro-heim-timing} shows how execution time scales with the number of trial DMs ($N_{\rm DM}$).
(1) \emph{Kernel scaling.} For both implementations the dedispersion kernel time grows approximately linearly with $N_{\rm DM}$, as expected for brute–force alignment. \textsc{Astroflow} remains substantially faster over the entire sweep: the kernel rises from $\sim$0.091\,s at $N_{\rm DM}=512$ to $\sim$1.295\,s at $8192$ (real–time factor $R\simeq221\rightarrow15.5$), whereas \texttt{Heimdall} increases from $\sim$1.21\,s at $N_{\rm DM}\!\approx\!543$ to $\sim$9.21\,s at $8298$ ($R\simeq16.5\rightarrow2.2$). 
(2) \emph{End-to-end time.} Including transfers, \textsc{Astroflow} grows from $\sim$0.26\,s ($N_{\rm DM}=512$) to $\sim$2.93\,s ($8192$), while \texttt{Heimdall} increases from $\sim$1.57\,s to $\sim$12.0\,s. Copy overhead is subdominant in both cases; for \textsc{Astroflow} it becomes comparable to the kernel at large $N_{\rm DM}$ (e.g., $\sim$1.63\,s copy vs.\ $\sim$1.30\,s kernel at $8192$), whereas \texttt{Heimdall} remains kernel–dominated. 
(3) \emph{Real–time performance.} The real–time factor $R$ decreases monotonically with $N_{\rm DM}$. \textsc{Astroflow} maintains faster–than–real–time operation ($R>1$) across the full range with ample margin; \texttt{Heimdall} also satisfies $R>1$ for these settings but with significantly lower margins. These trends reflect differences in kernel efficiency and the growing contribution of data movement at large $N_{\rm DM}$.

\subsection{End-to-End Runtime on Observational Data}
We evaluate the end-to-end wall time of \textsc{Astroflow} for single–pulse detection on real observations, excluding storage-to-host disk I/O because it is hardware dependent. The timing breakdown includes host–to–device and device–to–host memory copies (H2D/D2H), time downsampling, radio–frequency–interference (RFI) excision, dedispersion, host–side allocation of the dedispersed DM–time buffers, image gridding to form ${\rm DM}$–time slices, neural–network inference, and lightweight control/interop costs between C++ and Python (e.g., memory mapping via \texttt{pybind11}). The dataset is a contiguous $150$\,s observation with 8-bit quantization and native sampling $t_{\rm samp}=40\,\mu\mathrm{s}$; we search ${\rm DM}\in[0,1024]$ with $N_{\rm DM}=1024$ trials.

\begin{figure}
  \centering
  \includegraphics[width=0.95\linewidth]{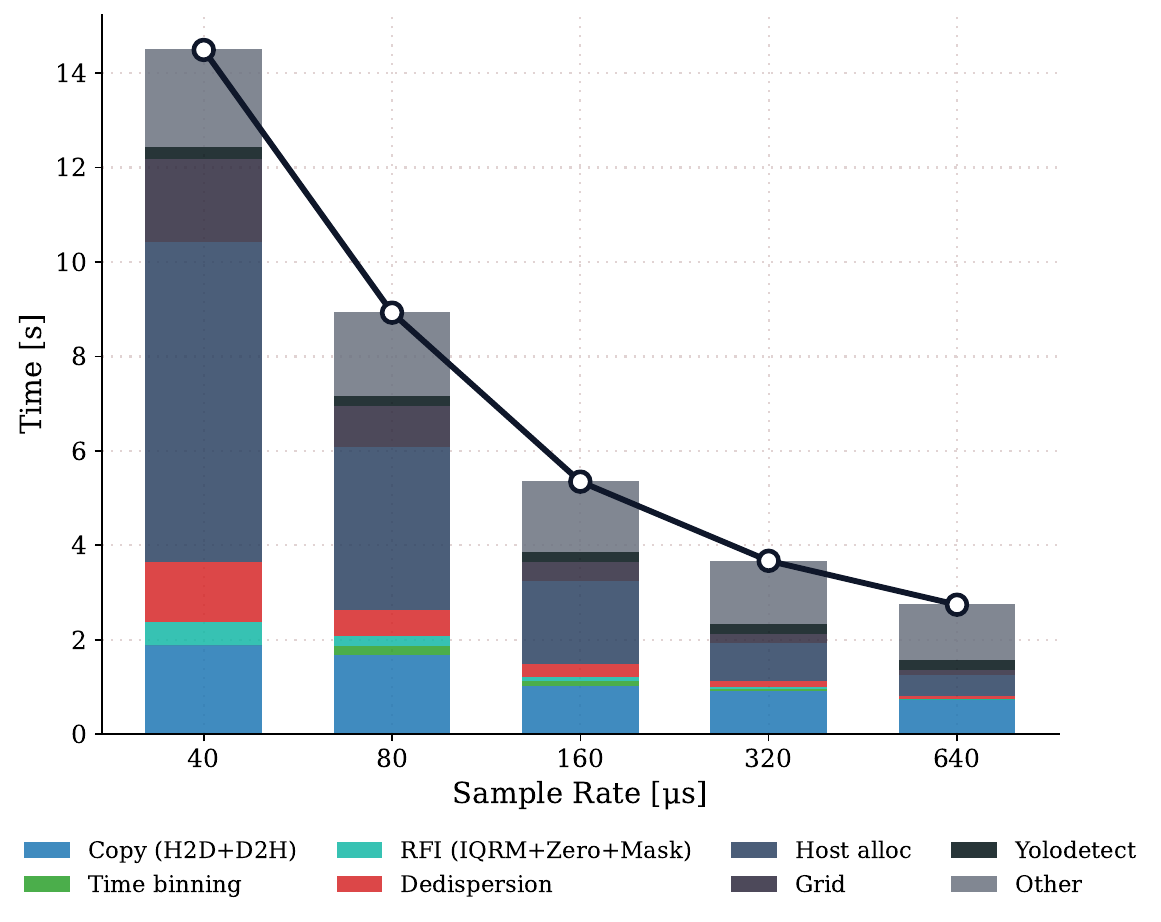}
  \caption{\textbf{End-to-end timing versus effective sample rate (downsampling factor).}
  Stacked bars report per–stage wall time on a $150$\,s observation (8-bit, native $t_{\rm samp}=40\,\mu$s, 
  $n_f=2048$, $f_c=1250$\,MHz, bandwidth $500$\,MHz) while searching ${\rm DM}\in[0,1024]$ with
  $N_{\rm DM}=1024$. Components: host-to-device and device-to-host copies
  (Copy; H2D+D2H), time downsampling (Time binning), RFI, dedispersion kernel (Stage~1+Stage~2),
  host-side allocation of DM–time buffers (Host alloc), image tiling/gridding to form
  ${\rm DM}$–time slices (Grid),
  YOLO inference time per batch (Yolodetect), and a small residual (“Other,” including control, checks, and minor frees).
  The dark line with white markers overlays the total runtime excluding disk I/O
  ($t_{\rm total}^{\rm no\text{-}IO}$); storage to host I/O is omitted because it is hardware dependent.}
  \label{fig:e2e-downsample}
\end{figure}

Figure~\ref{fig:e2e-downsample} shows that, once disk I/O is excluded, the total latency decreases monotonically as the effective sample rate is relaxed (i.e., with stronger downsampling), leaving comfortable faster-than-real-time headroom across the explored range. Because the input size is fixed, H2D transfer is effectively constant; by contrast, D2H shrinks with downsampling as the volume of dedispersed output scales with the number of effective samples, so the aggregated copy cost declines accordingly. GPU time binning and RFI excision remain lightweight owing to in-core streaming and locality-friendly kernels, and yield $\sim$20$\times$ speedups relative to an OpenMP CPU implementation under identical settings. Consistent with earlier discussion, the dedispersion kernel itself contributes only a minor fraction, indicating that the end-to-end workflow is predominantly \emph{memory–bound}. The dominant consumer at high time resolution is the host–side allocation of DM–time buffers together with subsequent tiling/gridding; both diminish as downsampling increases. Although these allocations could in principle be overlapped with computation, \textsc{Astroflow} has not yet implemented asynchronous management. Model inference scales with the number of tiles and is nearly constant in aggregate, averaging $\sim$0.5\,ms per image. The residual “Other’’ term reflects data validation, minor deallocation, and C++/Python interop (e.g., \texttt{pybind11}), and primarily scales with input size.

\section{Model Construction and Training}\label{sec:model}
This section describes the construction of the training set and the details of the training schedule. We assemble an FRB object–detection dataset of order $10^5$ images and apply diversified data augmentation to improve generalization. The training strategy combines backbone pretraining with staged fine–tuning, enabling efficient convergence and stable detection despite a compact parameter budget. The complete training procedure and the results on the $\sim10^5$–sample corpus are reported below; \textsc{Astroflow}'s performance on real observational fields is presented in Section~\ref{sec:real_obs}.

\subsection{Dataset Construction}
Object–detection supervision requires spatial annotations (location and scale) on images, whereas publicly available single–pulse resources remain incomplete. To cover a broader range of pulse morphologies and observing conditions, we compose the training set from the DRAFTS corpus \citep{DRAFTS2025}, the candidate set released by \citet{Agarwal2020}, and simulated pulses injected into observations from FAST \citep{FAST2011} and QUEST \citep{QUESTref}. QUEST (The Qilu University Explorer Survey Telescope) is a decimeter-wave array consisting of 20 on-axis dish antennas of 4.5-meter aperture. 

We additionally curate a large set of negatives composed of (i) background spectrograms from real observations and (ii) diverse RFI–contaminated frames. 
In real operating conditions, the spectral is typically \emph{coloured} and non–stationary and coexists with multiple forms of RFI \citep{Zhang2021coloured}. By injecting synthetic pulses directly into FAST/QUEST spectrograms for positives and sampling negatives from this mixture (background + RFI), then applying a unified dedispersion transform and gridding into pseudo–colour images, we construct a large, heterogeneous corpus that preserves real–world statistics and interference structure. This design enables the model to learn dispersion–coherent pulse signatures against coloured, non–stationary backgrounds and RFI.

\begin{figure}
    \centering
    \includegraphics[width=0.85\linewidth]{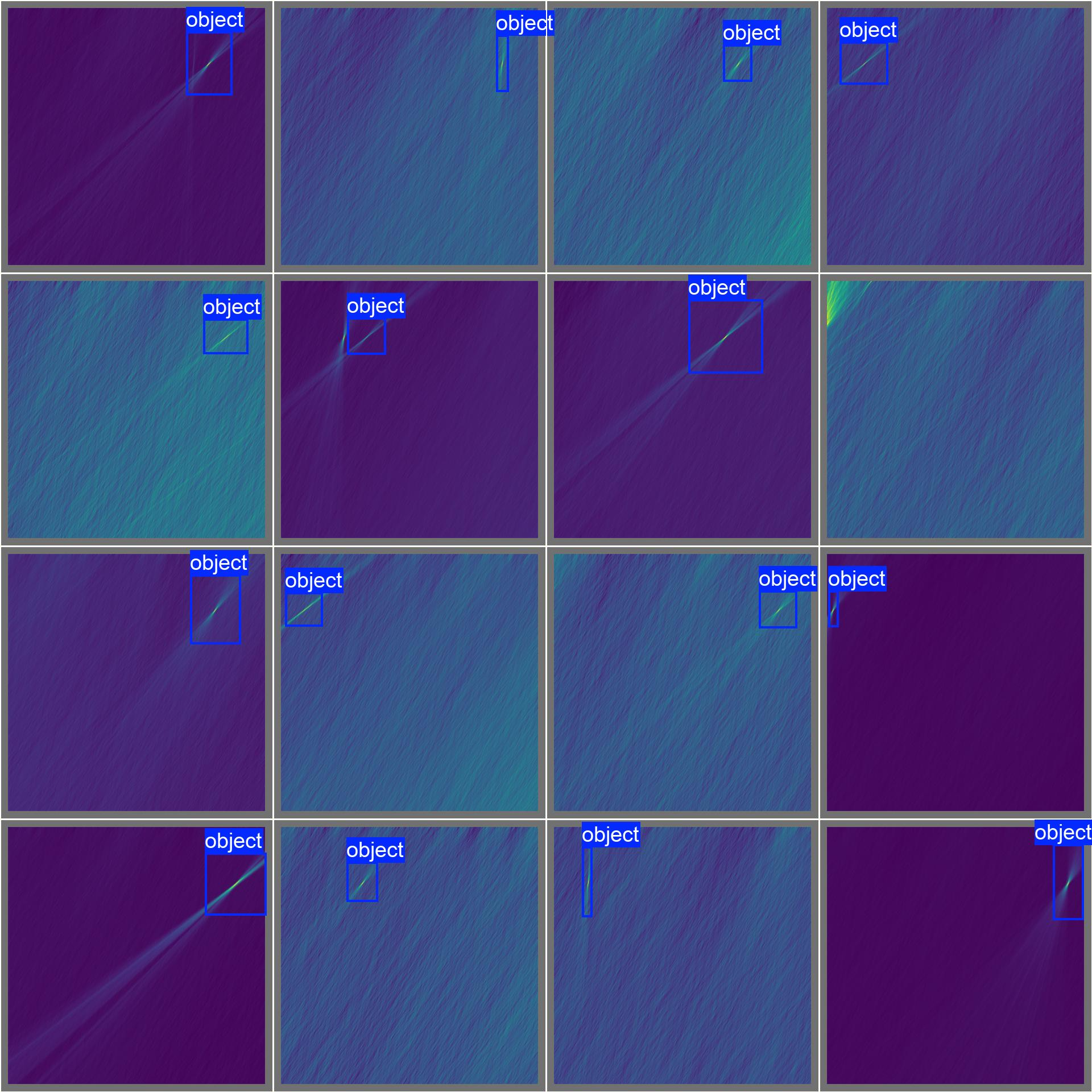}
    \caption{Representative annotation examples from the simulated single–pulse training set. 
    The blue bounding boxes highlight the injected pulse features in ${\rm DM}$–time (DMT) images, 
    which serve as the supervisory signal for object–detection training.}
    \label{fig:annotation}
\end{figure}

Given the scale of the corpus, exhaustive manual labeling is impractical; we therefore adopt a semi–automated scheme. Samples that already provide bounding boxes retain their original annotations; for candidate images whose ${\rm DM}$–time (DMT) pulse patterns are typically centered, we apply a uniform center annotation to maintain consistency; for injected samples, we use the known dispersion measure and time of arrival to generate programmatic bounding boxes. Summary statistics of the dataset are listed in Table~\ref{tab:dataset-split}, and representative annotation examples are shown in Figure~\ref{fig:annotation}.

\begin{deluxetable*}{l l r r r r r}
\tablecaption{Sample statistics of the constructed single–pulse detection dataset.\label{tab:dataset-split}}
\tablehead{
\colhead{Source} & \colhead{Reference} &
\colhead{Train (FRB)} & \colhead{Train (RFI)} &
\colhead{Test (FRB)}  & \colhead{Test (RFI)}  &
\colhead{Total}
}
\startdata
DRAFT & \citep{DRAFTS2025} & 1270  &    0 &  318  &    0 &  1588  \\
FETCH & \citep{Agarwal2020}      & 20000 & 20000 & 6664  &  7319 & 53983  \\
FAST  & —                  & 6552  & 8130  & 2186  &  3487 & 20355  \\
QUEST & —                  & 5737  & 16758 & 2895  &  8288 & 33678  \\
Total & —                  & 33559 & 44888 & 12063 & 19094 & 109604 \\
\enddata
\tablecomments{FRB: fast radio burst; RFI: radio–frequency interference.}
\end{deluxetable*}

\begin{figure}
    \centering
    \includegraphics[width=0.85\linewidth]{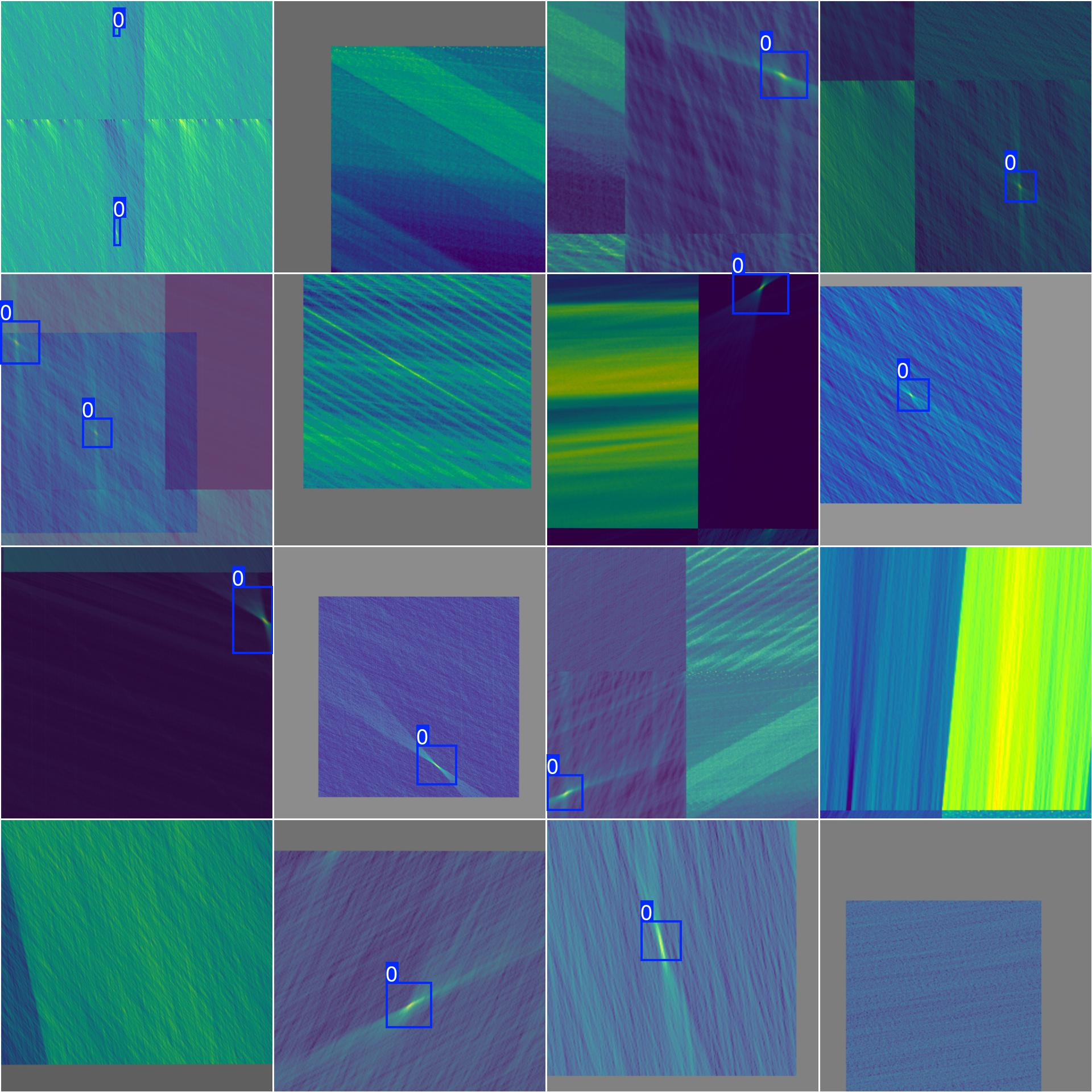}
    \caption{Examples of training samples after applying data–augmentation strategies. 
    Intensity, geometric, and contrast perturbations are introduced while preserving the underlying pulse morphology, 
    thereby improving the robustness of the detection model across diverse observing conditions and RFI environments.}
    \label{fig:enhannotation}
\end{figure}

To mitigate domain shift and enhance robustness while preserving the physical semantics of the pulses, we employ a suite of augmentations under a consistent rendering pipeline. Moderate perturbations are applied to intensity, geometry, and contrast; colour–space adjustments and local–contrast normalization are used to stabilize the visibility of fine–scale structures; and, without altering the pulse physics, diversified synthesis strategies are introduced to expand coverage of rare morphologies. After these treatments, the model exhibits more stable representations across telescopes, observing conditions, and RFI environments. Typical after examples appear in Figure~\ref{fig:enhannotation}, and the exact augmentation configuration and ranges are listed in Table~\ref{tab:aug_hparams_en}.

\subsection{Model Training}
Because the samples originate from multiple sources with heterogeneous shapes and scales, all inputs are first resampled to $512\times512$ to ensure a consistent rendering of the ${\rm DM}$--time geometry. In the candidate set of \citet{Agarwal2020}, annotations are centered by construction; as a result, even strong augmentation alone cannot fully remedy the tendency of the detector to overfit to the image center. In addition, the class distribution between single pulses and RFI is imbalanced, which makes a single-stage training recipe suboptimal for stable convergence and recall. Guided by these considerations, we adopt a pretraining--multi-stage fine-tuning schedule.

We begin with pretraining on $\sim$20k simulated FRB samples from \textsc{FETCH}, which primes the network to capture pulse morphology and characteristic local structures in ${\rm DM}$--time slices \citep{Agarwal2020}. After this warm start, we remove all center-labeled samples (including both FRB and RFI backgrounds) and retrain the model with a smaller learning rate, encouraging sensitivity to global position rather than the central region alone. During the subsequent phase, to alleviate the positive/negative imbalance, we resample positives in a controlled manner so that the detector receives a more informative mix of examples across training iterations. Finally, we freeze part of the backbone and fine-tune the detection head to stabilize localization and classification, followed by an unfrozen, global fine-tuning pass that consolidates the improvements. This procedure yields the final model used in our experiments. 

\subsection{Evaluation Metrics}
We evaluate the detector with standard criteria, aiming to accurately identify single pulses while minimizing the probability of misclassifying radio frequency interference (RFI) as candidates. Specifically, we report accuracy, precision, recall, and the F$_1$ score. Accuracy is the ratio of correct predictions (FRB and RFI) to the total number of predictions. Precision is the fraction of correctly labeled FRBs among all instances predicted as FRB. Recall is the fraction of true FRBs that are correctly identified. The F$_1$ score is the harmonic mean of precision and recall, providing a balanced summary of the two. During computation, single pulses from FRBs and pulsars are treated as the positive (true) class, whereas RFI is treated as the negative class.

\subsection{Results}
\begin{figure*}[htbp]
    \centering
    \begin{minipage}[b]{0.48\linewidth}
        \centering
        \includegraphics[width=\linewidth]{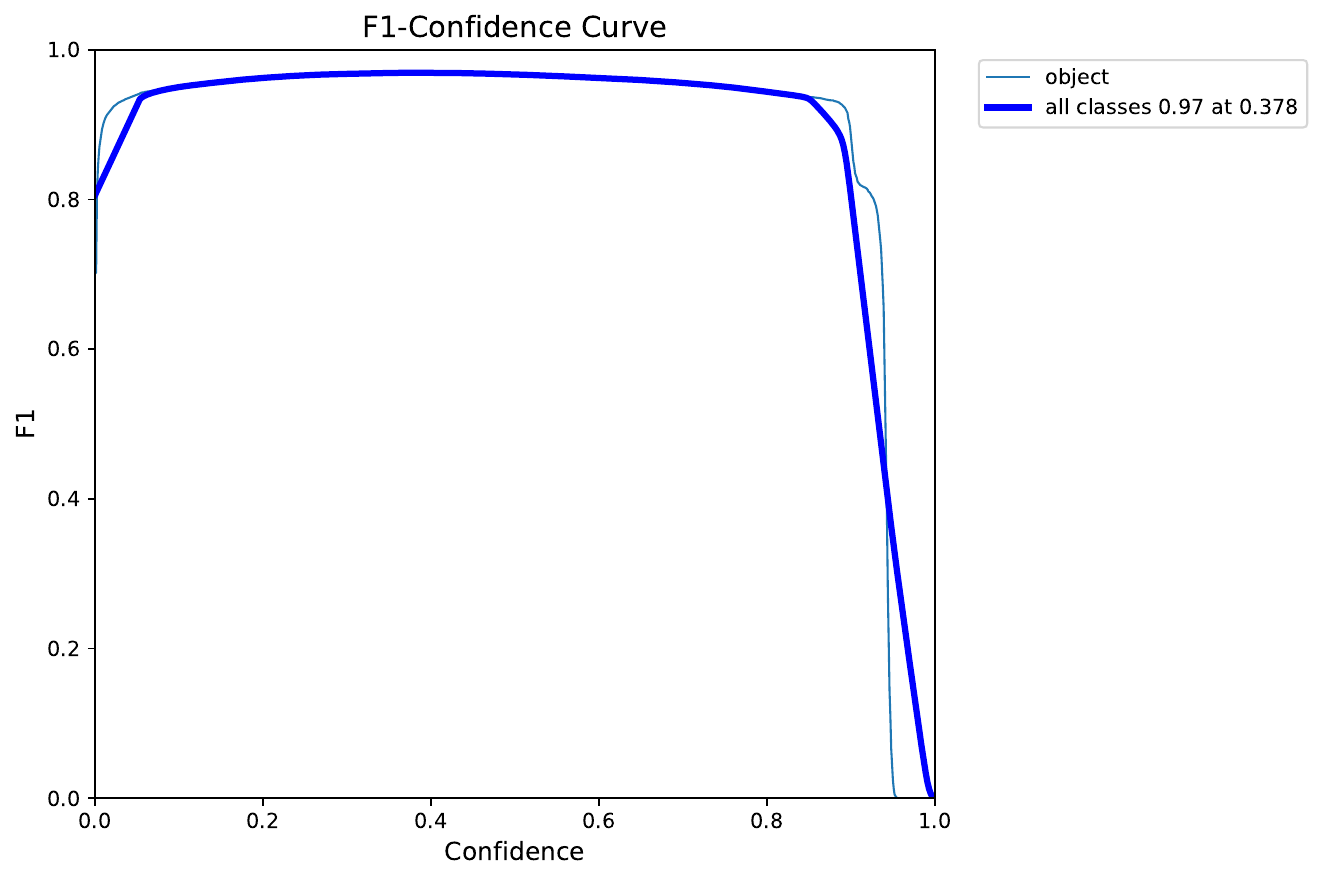}
    \end{minipage}
    \hfill
    \begin{minipage}[b]{0.48\linewidth}
        \centering
        \includegraphics[width=\linewidth]{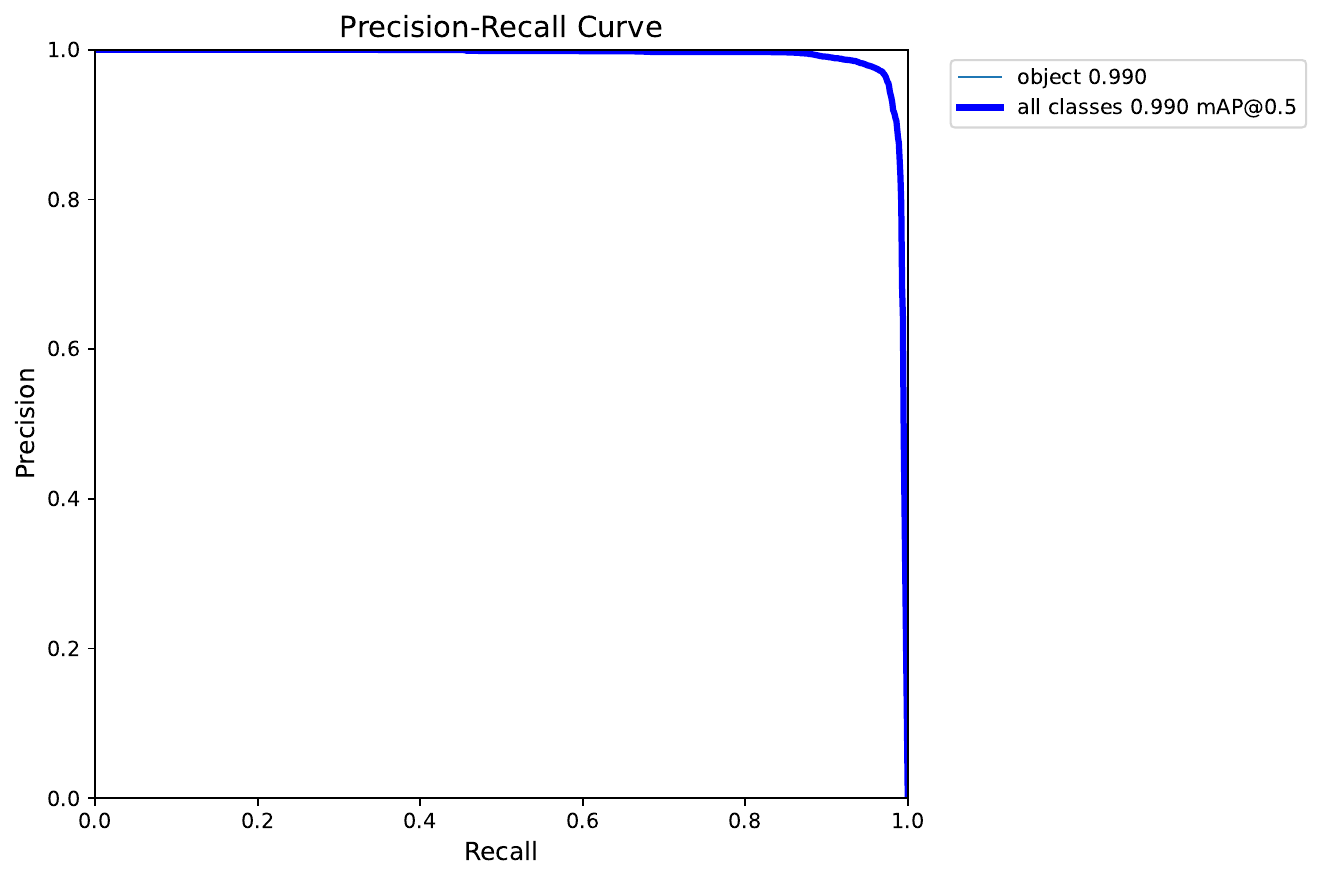}
    \end{minipage}
    \caption{\textbf{Validation curves.} \emph{Left:} F$_1$--confidence curve, peaking near $0.97$ at a confidence threshold of $\sim$0.378. 
    \emph{Right:} Precision--recall (PR) curve with precision $\approx 1$ over most of the recall range; the mAP@0.5 reaches $0.990$.}
    \label{fig:f1-pr-curve}
\end{figure*}

Figures~\ref{fig:f1-pr-curve} summarizes the performance on the validation set. 
The precision--recall (PR) curve indicates that precision remains close to 1 across almost the entire recall range, with mAP@0.5 reaching 0.990. 
The F$_1$--confidence curve shows that, at an appropriate confidence threshold (approximately 0.378), the F$_1$ score approaches 0.97, reflecting a favorable balance between precision and recall. 
Together, these results suggest that the  \texttt{YOLOv11N} based detector achieves strong generalization and robustness on the constructed dataset, effectively limiting RFI false positives while maintaining high accuracy.

\section{Results on Observations}\label{sec:real_obs}
To assess both the detection efficiency and the false–alarm behavior of \textsc{Astroflow} under real telescope operating conditions, we analyze two complementary data sets. 
First, we use the independent FAST–FREX corpus \citep{FAST_FREX2024}, which comprises $600$ burst samples drawn from several FRB sources; each event is provided as a $\simeq6.04$\,s FITS file together with the release’s best–estimate dispersion measure. 
Second, we report a deployment on a compact 4.5\,m array (four antennas; L band 1000–1500\,MHz, $N_{\rm chan}=512$), where \textsc{Astroflow} was first validated by detecting giant pulses from the Crab pulsar and then operated in a ten–day, continuous FRB survey mode.

In the survey configuration—filterbank spectra with a time resolution of 196\,$\mu$s and 512 frequency channels—the pipeline sustained $\sim$$10\times$ real–time throughput on a single host and processed a total exposure of $\sim4\times24\times10\approx960$ antenna–hours, yielding 4135 candidates with no significant FRB detection. 
The following subsections detail the configurations and quantitative results for both data sets.

\subsection{FAST\_FREX}\label{subsec:fast_frex}

According to the documentation, each FAST\_FREX file spans $6.04$\,s, consists of $4096$ frequency channels, and has a native sampling time of either $40\,\mu$s or $96\,\mu$s. 
The dataset contains bursts from multiple repeating FRB sources, including FRB\,20121102, FRB\,20180301, and FRB\,20201124, with a total of $600$ samples. 
Each file provides a reference $(\mathrm{DM}, \mathrm{TOA})$ pair, and a detection is regarded as correct if it corresponds to the same burst event.

For benchmarking, we evaluated \textsc{Astroflow} against two widely adopted single–pulse search pipelines: \texttt{Heimdall} and \texttt{transientX} \citep{transientx2024}. 
\textsc{Heimdall} is a well–established GPU–based single–pulse search framework, while \textsc{TransientX} is an optimized CPU–based software that integrates RFI mitigation, dedispersion, matched filtering, and candidate clustering. 
It employs a DBSCAN–based algorithm to consolidate detections, providing efficient FRB search capabilities.

Since \textsc{Heimdall} does not natively support PSRFITS input, we used a lightweight wrapper (\texttt{your\_heimdall.py}) that directly interfaces the PSRFITS data stream with the \textsc{Heimdall} processing core without data conversion\citep{your2020}. 
The reported runtime includes only the execution of the main \textsc{Heimdall} process. 
All searches adopted a common DM range of $100$–$700$\,pc\,cm$^{-3}$ and a frequency band of 1000–1500\,MHz.

A detection is classified as a \emph{true positive (TP)} if it satisfies $|\Delta\mathrm{DM}| < 20$ and $|\Delta\mathrm{TOA}| < 0.2$\,s, referenced to 1500\,MHz. 
For \textsc{Heimdall} and \textsc{TransientX}, the TOA tolerance is relaxed to $0.4$\,s to account for repeated detections of the same burst. 
True positives are de–duplicated across parameter settings, while all unmatched detections are counted as \emph{false positives (FP)}. 
All TP events were manually verified, whereas FPs were not manually inspected due to their large number.

\begin{figure}
  \centering
  \includegraphics[width=0.98\linewidth]{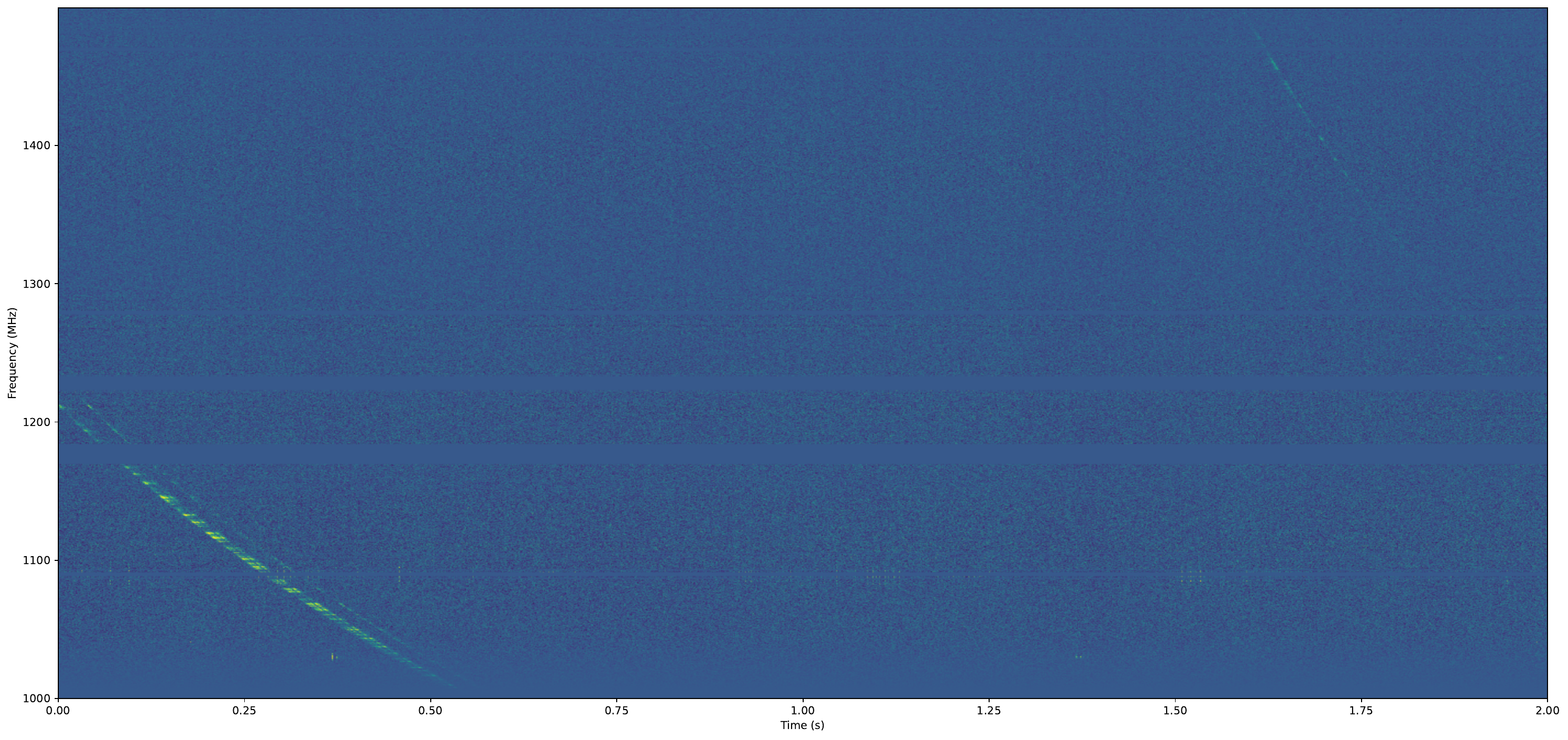}
  \caption{Example FAST\_FREX file with two bursts in the same record.
  Dynamic spectrum over 1.0–1.5\,GHz showing only the first 2.0\,s of the 6.04\,s file. 
  Two temporally separated bursts are visible in this excerpt. 
  The release provides a single reference $(\mathrm{DM},\mathrm{TOA})$ for the file; the additional, unannotated burst is \emph{excluded from scoring} for all methods (neither TP nor FP), although it is detected by \textsc{Astroflow}.}
  \label{fig:two_bursts_in_one_file}
\end{figure}

\begin{figure}
  \centering
  \includegraphics[width=0.98\linewidth]{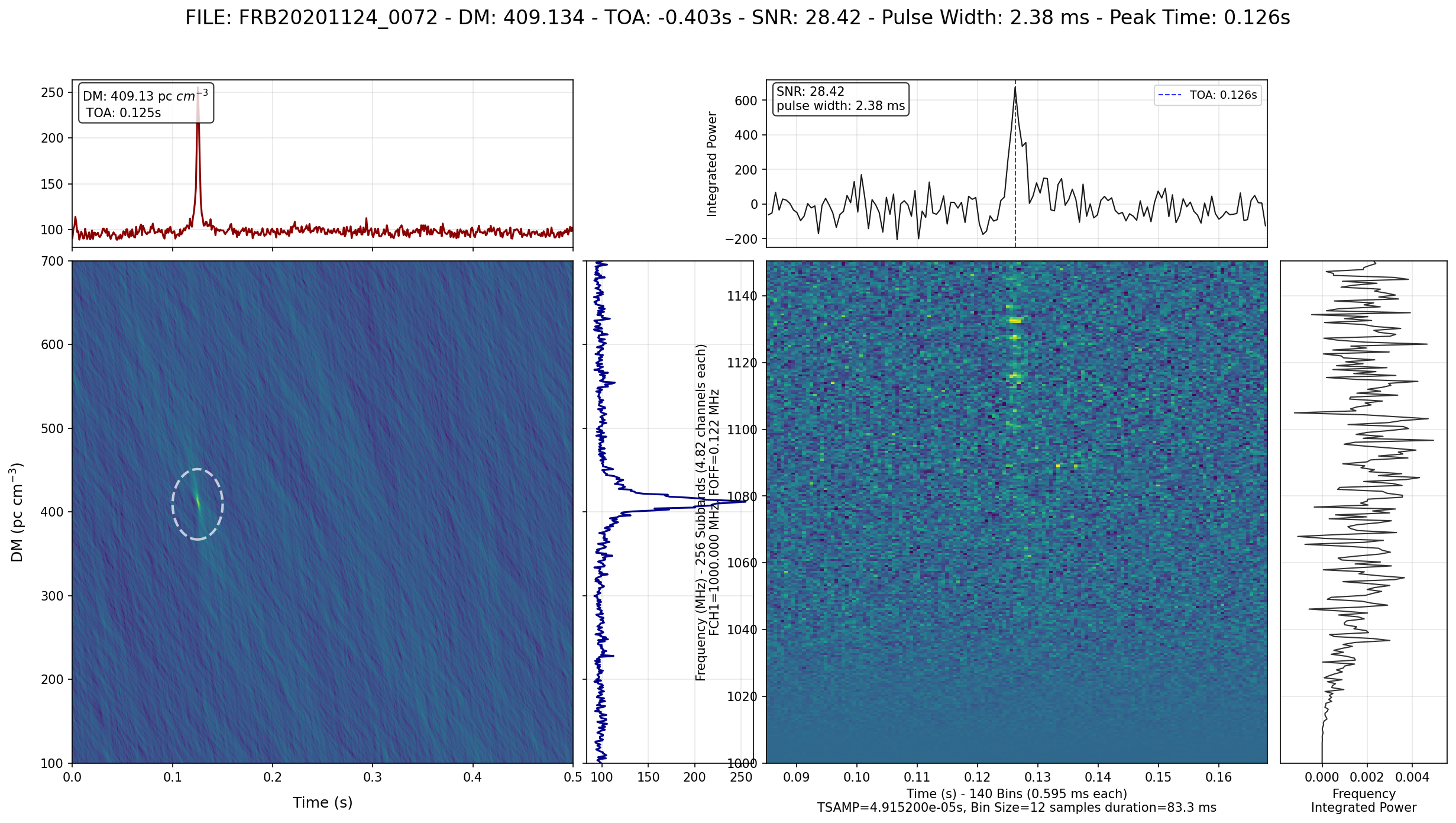}
  \caption{\textsc{Astroflow} candidate from an unannotated pulse: DM–matched, TOA–mismatched.
  Quick–look panels show the DM–time map and the dedispersed time series at the best–fit DM.
  The best–fit DM agrees with the catalog DM (within $|\Delta\mathrm{DM}|<20$), but the detected TOA differs from the catalog TOA by more than the adopted tolerance; hence, this event is excluded from scoring (neither TP nor FP).}
  \label{fig:dm_match_toa_mismatch}
\end{figure}

We found that a subset of FAST\_FREX files actually contains more than one astrophysical pulse, even though the release provides only a single reference $(\mathrm{DM},\mathrm{TOA})$ per file and does not annotate additional pulses.
Figure~\ref{fig:two_bursts_in_one_file} shows a representative example: two distinct dispersion tracks are present in the same 6.04\,s file, but only one is catalog–annotated.
\emph{For consistency with the ground truth}, detections associated with such unannotated pulses are excluded from scoring for all methods (neither TP nor FP), although \textsc{Astroflow} is able to detect them.
For clarity, Fig.~\ref{fig:dm_match_toa_mismatch} illustrates a typical \textsc{Astroflow} candidate from an unannotated pulse that matches the catalog DM but fails the TOA tolerance; it is therefore not counted.

Table~\ref{tab:configs} summarizes the \textsc{Astroflow} configurations used on the FAST\_FREX dataset. 
The three variants (A1–A3) share the same time grid ($T_{\rm grid}=0.5$\,s), DM range ($100$–$700$\,pc\,cm$^{-3}$), and time downsampling factor (8), differing only in their bandwidth combinations: A1 uses the full band (F; 1000–1500\,MHz), while A2 and A3 use multi–band combinations (F+L+H). 
The column \texttt{conf} denotes the classification confidence threshold for the deep–learning stage.

\begin{deluxetable*}{llccccc}
\tablewidth{\textwidth}
\tablecaption{Search configurations used on the FAST\_FREX dataset.\label{tab:configs}}
\tablehead{
\colhead{No.} & \colhead{Software} & \colhead{conf} &
\colhead{$T_{\rm grid}$ (s)} & \colhead{DM range (pc\,cm$^{-3}$)} &
\colhead{Bandwidth} & \colhead{Time downsampling}
}
\startdata
A1 & \textsc{Astroflow} & 0.4 & 0.5 & 100–700 & F       & 8  \\
A2 & \textsc{Astroflow} & 0.5 & 0.5 & 100–700 & F+L+H   & 8  \\
A3 & \textsc{Astroflow} & 0.4 & 0.5 & 100–700 & F+L+H   & 8  \\
\enddata
\tablecomments{F $\equiv$ 1000–1500\,MHz; L $\equiv$ 1000–1250\,MHz; H $\equiv$ 1250–1500\,MHz.}
\end{deluxetable*}

The configurations of \textsc{Heimdall} and \textsc{TransientX} are summarized in Table~\ref{tab:configs_baselines}. 
Since their internal parameter structures differ from \textsc{Astroflow}, only the key runtime parameters are listed, including the DM range, RFI settings, and time–downsampling schemes. 

\begin{deluxetable*}{llccc}
\tablewidth{\textwidth}
\tablecaption{Key runtime parameters for \textsc{Heimdall} and \textsc{TransientX}.\label{tab:configs_baselines}}
\tablehead{
\colhead{No.} & \colhead{Software} &
\colhead{DM range (pc\,cm$^{-3}$)} & \colhead{Time downsampling} & \colhead{RFI setting}
}
\startdata
H1 & \textsc{Heimdall}   & 100–700 & adaptive & \texttt{rfi\_tol}=5,\ baseline=2 \\
T1 & \textsc{TransientX} & 100–700 & 8 & \texttt{zdot kadaneF(8,4)} \\
\enddata
\tablecomments{Both pipelines were executed over 1000–1500\,MHz. 
The ``adaptive'' mode in \textsc{Heimdall} denotes its internal time–scrunching strategy during dedispersion.
\textsc{TransientX} was run on 32 CPU threads with DBSCAN clustering enabled.}
\end{deluxetable*}

Table~\ref{tab:results} presents the detection statistics following the unified matching and counting criteria. 
\textsc{Astroflow} achieves high recall while maintaining extremely low false–positive rates, resulting in a precision above 95\%. 
In contrast, \textsc{Heimdall} and \textsc{TransientX} successfully recover most astrophysical bursts but produce substantially more false detections. 
The per–iteration runtimes are also listed; on the FAST\_FREX benchmark, \textsc{Astroflow} outperforms both baselines in speed and reliability.

\begin{deluxetable*}{llllllllll}
\tablewidth{\textwidth}
\tablecaption{Summary of detection results on FAST\_FREX.\label{tab:results}}
\tablehead{
\colhead{No.} & \colhead{Software} & \colhead{TP} & \colhead{FP} &
\colhead{P (\%)} & \colhead{R (\%)} & \colhead{Total} &
\colhead{Speed [s/it]} & \colhead{Bandwidth} & \colhead{Confidence}
}
\startdata
H1 & \textsc{Heimdall}    & 487 & 8635  & 5.33 & 81.2 & 600 & 4.72 & F     & – \\
T1 & \textsc{TransientX}  & 576 & 10044 & 5.42 & 96.0 & 600 & 3.13 & F     & – \\
A1 & \textsc{Astroflow}   & 571 & 3     & 99.5 & 95.2 & 600 & 0.25 & F     & 0.4 \\
A2 & \textsc{Astroflow}   & 584 & 28    & 95.4 & 97.3 & 600 & 0.67 & F+L+H & 0.5 \\
A3 & \textsc{Astroflow}   & 588 & 49    & 92.3 & 98.0 & 600 & 0.65 & F+L+H & 0.4 \\
\enddata
\tablecomments{P: precision; R: recall. Bandwidth labels follow Table~\ref{tab:configs}.}
\end{deluxetable*}

\subsection{Real Observations with the QUEST 4.5\,m Array}

The QUEST facility, located in Shandong Province, China, is an under-construction array of twenty 4.5\,m radio dishes operating in the L band (1.05–1.45\,GHz). 
The system is currently in the commissioning stage.
To evaluate the performance of \textsc{Astroflow} on real telescope data, we selected four antennas to conduct FRB search experiments.
The data flow proceeds as follows: baseband voltages are channelized online into filterbank products ($N_{\rm chan}=512$), which are continuously monitored by \textsc{Astroflow} for real-time dedispersion and candidate generation.

\begin{figure}
\centering
\includegraphics[width=0.98\linewidth]{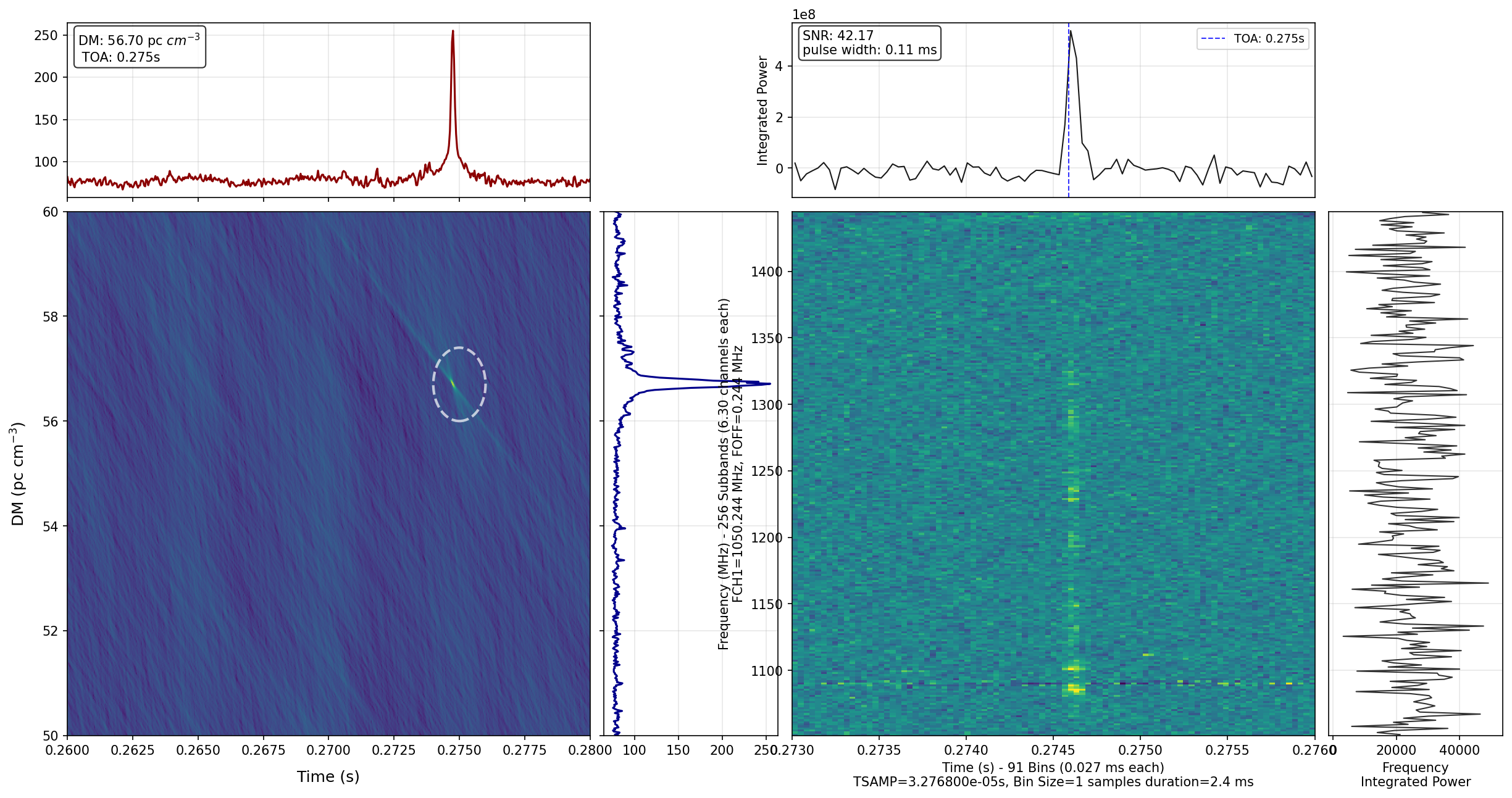}
\caption{Typical Crab giant–pulse detection with the QUEST 4.5\,m array. 
Left: DM–time map highlighting the dispersion track at the $\mathrm{DM}=56.7~\mathrm{pc\,cm^{-3}}$; 
right: dedispersed dynamic spectrum and summed profiles across 1.0–1.5\,GHz.}
\label{fig:crab_gp}
\end{figure}

As an end-to-end validation of system connectivity, we conducted short observations of the Crab pulsar and used the detection of giant pulses (GPs) as a functional test. 
Given the intrinsically narrow widths of Crab GPs, the validation run employed a 40\,$\mu$s time resolution and 512 frequency channels. 
Multiple GPs were successfully detected, and the resulting dynamic spectra and dedispersed time series exhibited the expected morphology, confirming the correctness of the entire signal chain—from the antennas through GPU processing to candidate output.
A representative output is shown in Fig.~\ref{fig:crab_gp}.

A ten-day continuous survey targeting millisecond-scale FRBs was subsequently performed.
In this mode, the filterbank configuration used a 129\,$\mu$s sampling interval and 512 frequency channels.
Data from four antennas were transmitted via a high-speed link to a processing server, where a single host executed \textsc{Astroflow} to search all antenna streams in real time.
The server hardware comprised two Intel(R) Xeon(R) Silver 4314 CPUs and one NVIDIA RTX 4090 GPU.
The system operated continuously for approximately 10 days ($24$\,h/day), corresponding to a total on-sky time of $4\times24\times10\approx960$ antenna-hours.
Under this configuration, \textsc{Astroflow} achieved a processing throughput of about 10$\times$ real time without any backlog.

The survey covered three DM sub-ranges: 
10–100, 100–800, and 800–1600\,pc\,cm$^{-3}$, 
each searched with a time grid of $T_{\rm grid}=0.5$\,s, 
a time downsampling factor of 1, and an IQRM mask for narrowband RFI mitigation. 
All runs used the full 1000–1500\,MHz bandwidth. 
The confidence threshold (\texttt{conf}) for the deep-learning classifier was set to 0.5.

During the ten-day continuous survey, \textsc{Astroflow} detected a total of 4135 candidates, with no significant FRB events identified.
Normalized by total exposure time, this corresponds to an average trigger rate of $\sim$4.3 per antenna-hour.
Overall, the false trigger rate remained modest, and the candidate volume was manageable for real-time review and post-processing.

\begin{figure*}[htbp!]
\centering
\gridline{
  \fig{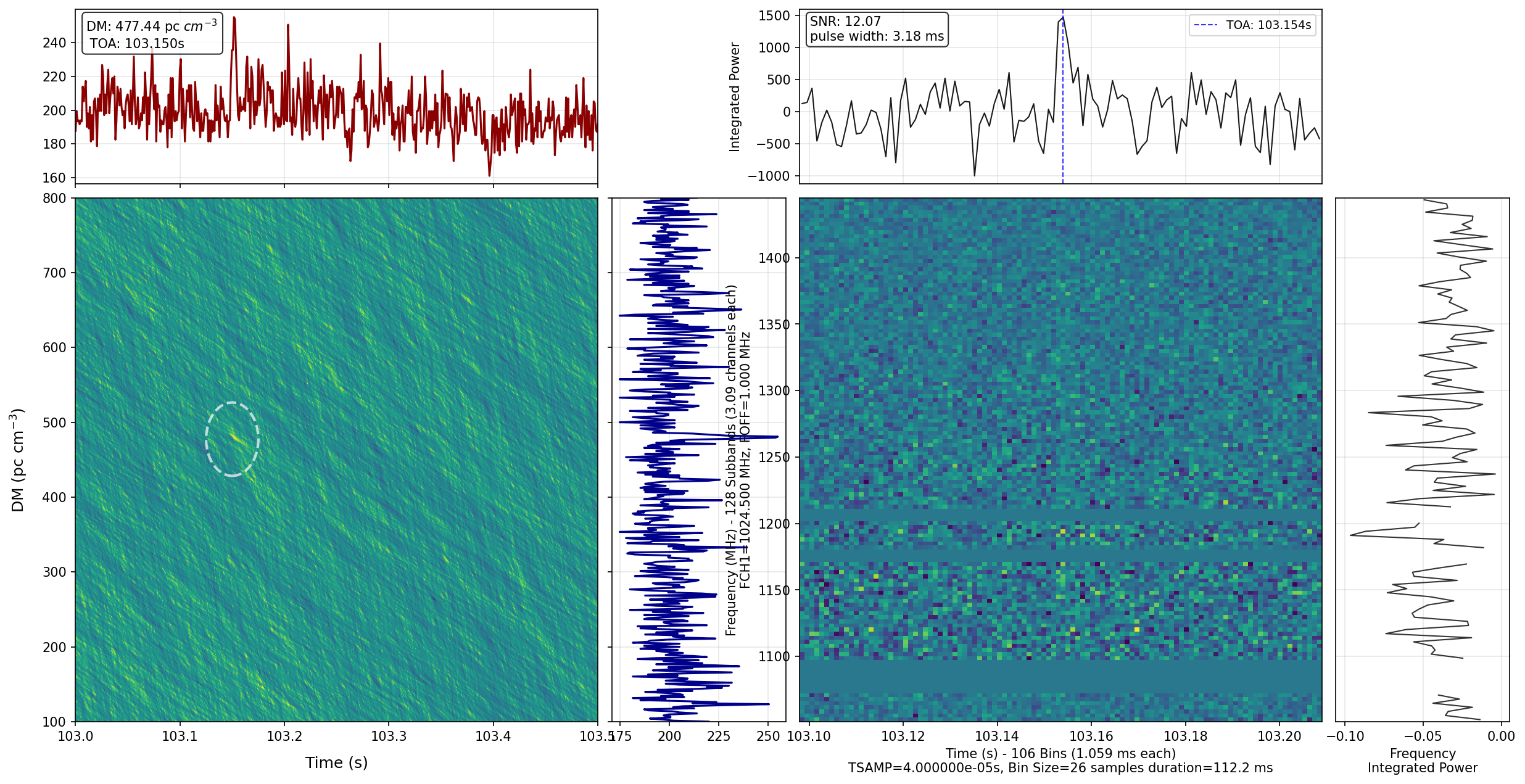}{0.32\textwidth}{(a) Burst-like fluctuations}
  \fig{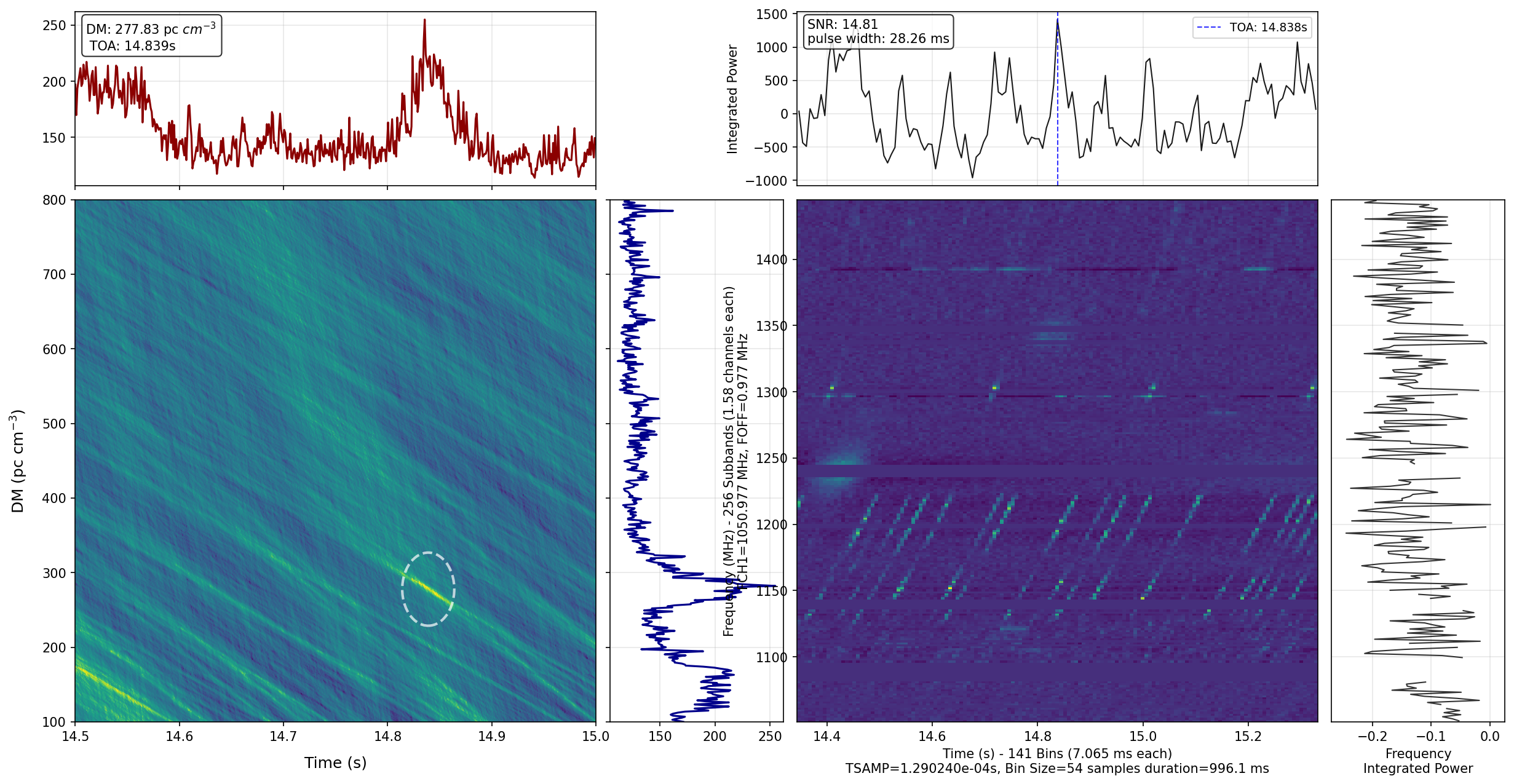}{0.32\textwidth}{(b) Impulsive broadband RFI}
  \fig{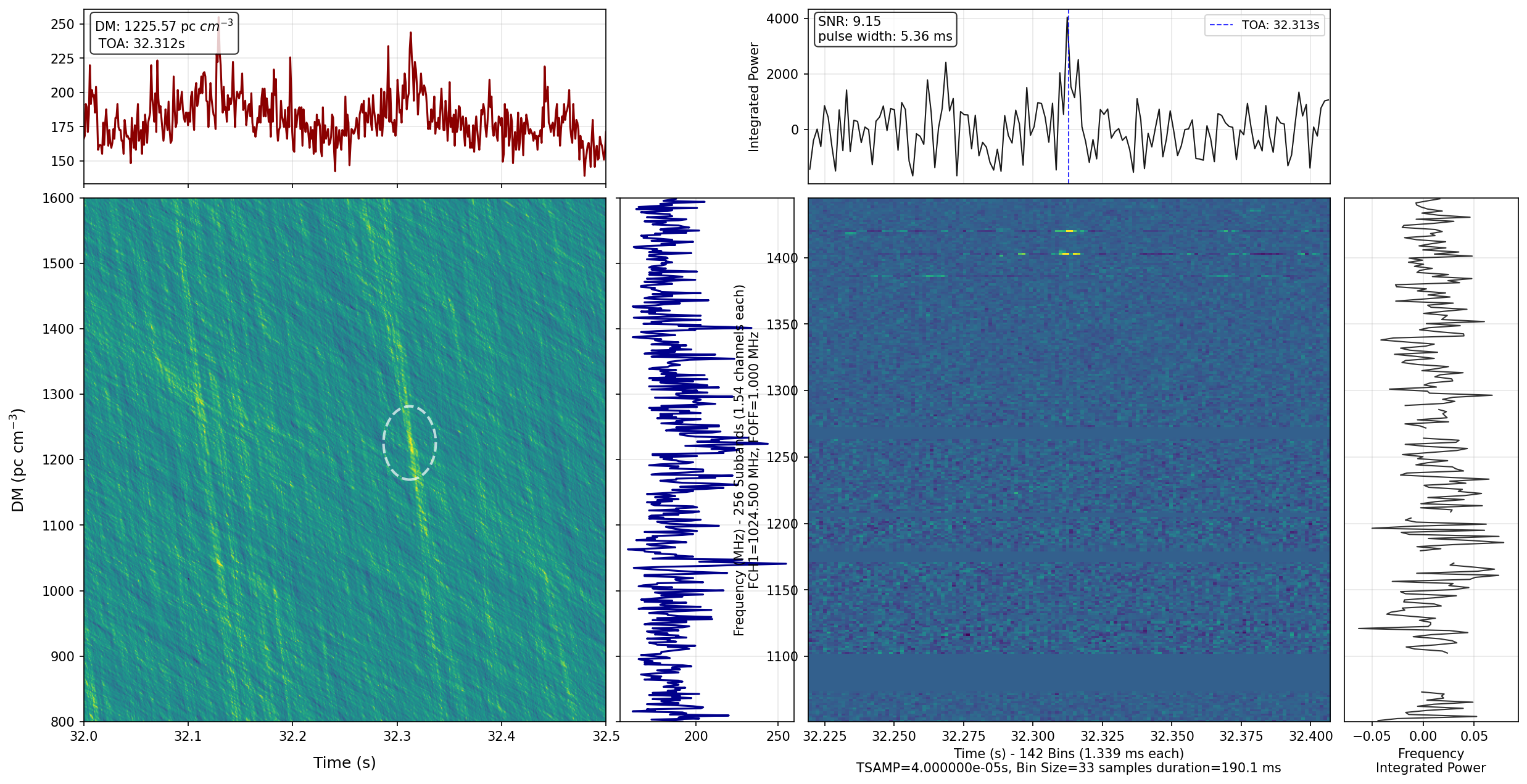}{0.32\textwidth}{(c) Narrowband-dominated noise}
}
\caption{
Examples of non-astrophysical triggers from the QUEST run, illustrating the three dominant classes discussed in the text. 
\textbf{(a) Burst-like fluctuations}: clustered tracks in the DM–time map that mimic an FRB morphology, yet the dedispersed dynamic spectrum reduces to noise-level structure without a coherent broadband spike. 
\textbf{(b) Impulsive broadband RFI}: near-bandwide, short-duration impulses that are more likely to pass IQRM (which primarily targets narrowband features); note the broadband excess in the dedispersed spectrum. The zero-DM filter (supported by \textsc{Astroflow}) was disabled in this survey to avoid potential suppression of weak bursts and will be considered in future refinements. 
\textbf{(c) Narrowband-dominated noise}: power concentrated at a few fixed frequencies over a short time interval; these events may form clusters in the DM–time plane but collapse to frequency-localized fluctuations in the dedispersed spectrum rather than a coherent broadband pulse.
All panels are generated by the same quick-look tool and share the 1.0–1.5\,GHz band with the indicated time sampling and channelization.}
\label{fig:rfi_gallery}
\end{figure*}

Morphologically, non-astrophysical triggers (RFI or false detections) can be broadly grouped into three categories(see Figure~\ref{fig:rfi_gallery}):
(1) strong impulsive broadband RFI—these are more likely to pass the IQRM filter, which primarily targets narrowband interference; although \textsc{Astroflow} supports zero-DM filtering, it was not enabled in this run to avoid excessive suppression of weak bursts and will be considered in future upgrades;
(2) burst-like fluctuations—features that form apparent clustered structures in the DM–time plane but reduce to noise-level variations in dedispersed spectra; and
(3) \emph{narrowband-dominated noise bursts}—power concentrated at a few fixed frequencies over short intervals, which can mimic DM–time clustering while remaining frequency-localized after dedispersion.

These false triggers can be further reduced through improved threshold tuning, enhanced RFI mitigation, and the incorporation of zero-DM filtering in future pipeline iterations.

The results demonstrate that \textsc{Astroflow} can stably perform long-duration, real-time FRB searches on the QUEST 4.5\,m array.
Running on a single host, the pipeline achieved $\sim$10$\times$ real-time throughput while maintaining a low false-trigger rate.
The successful detection of Crab giant pulses verifies the correctness of the full end-to-end system.
Future work will focus on improving RFI suppression and fine-tuning the deep-learning model to further reduce false positives and enhance genuine burst recovery.

\section{Discussion}\label{sec:discussion}

\subsection{End-to-End Performance and System Factors}
The benchmarks in Sec.~\ref{sec:benchmarking} show that end-to-end latency is shaped more by memory transfers and host-side buffer allocation than by the dedispersion kernel itself. storage-to-host disk I/O strongly depends on the medium (HDD vs.\ SSD/NVMe) and was excluded from the core timing. Once data are in host memory, host-to-device (H2D) transfers are constant for a fixed input size, whereas device-to-host (D2H) decreases with downsampling because the dedispersed output volume scales with the effective sample count. At high time resolution, host buffer allocation and tiling dominate, confirming that the workflow is largely memory-bound. Asynchronous allocation could in principle overlap with GPU kernels, but this optimization is not yet implemented.

\subsection{Image-Based Detection and Parameter Dependence}
Since the detector operates on ${\rm DM}$--time images, the observable morphology depends on search parameters. Very narrow pulses ($W/T_{\rm grid}<0.02$) compress the bow--tie feature and are more likely to be missed, while too narrow a \texttt{dmrange} or underspecified \texttt{dmtrail} induces stretching or distortion along the DM axis. These effects arise from parameterization rather than the detector itself. In practice, a small set of complementary configurations (e.g., pairing a nominal setting with one optimized for narrow pulses) and taking the union of detections provides more complete coverage.

\subsection{Fine-Tuning and Iterative Improvement}
On an RTX~4090, fine-tuning the YOLOv11N-based model with additional samples can be completed within $\sim$30 minutes for a $10^5$-sample corpus. This enables iterative refinement: newly detected pulses or improved injection schemes can be recycled into training, gradually improving robustness to rare morphologies and complex RFI backgrounds without retraining from scratch.

\subsection{Availability and Usability}
\textsc{Astroflow} is released as a Python package on PyPI and can be installed via \texttt{pip install pulseflow}. A YAML-based command-line interface facilitates deployment in standard pipelines, and extensive documentation assists user onboarding.\footnote{\url{https://github.com/lintian233/astroflow}} These features lower the barrier for integrating real-time searches into diverse observational facilities.

\section{Conclusion}\label{sec:conclusion}
Our study demonstrates that \textsc{Astroflow} achieves both high accuracy and faster-than-real-time throughput for single-pulse searches. On the validation set, the  \texttt{YOLOv11N} detector reaches mAP@0.5 of 0.990 and an F$_1$ score of $\sim$0.97 at the operating threshold. On FAST\_FREX, \textsc{Astroflow} recovers 95--98\% of astrophysical bursts with precision above 92\% and per-iteration latency below 1\,s on a single RTX~4090 GPU.

Combined with a GPU-optimized dedispersion kernel, robust RFI excision, and a lightweight image-based detection module, \textsc{Astroflow} provides an end-to-end, easy-to-deploy solution for real-time searches. Its accuracy, efficiency, and usability make it well suited for upcoming large-scale time-domain surveys where completeness and operational performance are both critical. It has already been deployed for the QUEST survey and will support more wide-field surveys in the future.

\begin{acknowledgments}
This work was supported by the National Natural Science Foundation of China (NSFC) grant No. 12375108. Jie Zhang is supported by National Key R\&D Program of China (2023YFB4503305) and the National Natural Science Foundation of China (Grants No. 12373109).
\end{acknowledgments}

\newpage
\appendix

\begin{table}[htbp]
\centering
\caption{Data augmentation hyperparameters used in \texttt{YOLOv11n} training. Values follow the Ultralytics YOLOv11 default configuration with slight adjustments.}
\begin{tabular}{lll}
\toprule
Parameter & Description & Value \\
\midrule
\texttt{hsv\_h} & Hue variation & 0.015 \\
\texttt{hsv\_s} & Saturation variation & 0.30 \\
\texttt{hsv\_v} & Value variation & 0.40 \\
\texttt{degrees} & Rotation (degrees) & 0.2 \\
\texttt{translate} & Translation fraction & 0.20 \\
\texttt{scale} & Scaling factor & 0.30 \\
\texttt{fliplr} & Horizontal flip probability & 0.5 \\
\texttt{mosaic} & Mosaic probability & 0.35 \\
\texttt{mixup} & MixUp probability & 0.12 \\
\texttt{copy\_paste} & Copy–paste probability & 0.20 \\
\texttt{copy\_paste\_mode} & Copy–paste mode & \texttt{mixup} \\
\texttt{auto\_augment} & Auto augmentation & \texttt{autoaugment} \\
\bottomrule
\end{tabular}
\label{tab:aug_hparams_en}
\end{table}


\bibliography{sample701}{}
\bibliographystyle{aasjournalv7}



\end{document}